\documentclass[%
 reprint,
 amsmath,amssymb,
 aps,
 superscriptaddress,
]{revtex4-1}
\usepackage{csquotes}
\usepackage{ulem}
\usepackage{overpic}
\usepackage{graphicx}
\usepackage{dcolumn}
\usepackage{bm}
\usepackage{hyperref}
\usepackage{xcolor}
\usepackage{braket}
\usepackage{changes}

\begin{document}

\preprint{APS/123-QED}

\title{Thermodynamic precision in the nonequilibrium exchange scenario}
\author{Donato Farina}
\email{donato.farina@icfo.eu}
\affiliation{ICFO - Institut de Ciencies Fotoniques, The Barcelona Institute of Science and Technology, Castelldefels (Barcelona) 08860, Spain}

\author{Bilal Benazout}
\affiliation{ICFO - Institut de Ciencies Fotoniques, The Barcelona Institute of Science and Technology, Castelldefels (Barcelona) 08860, Spain}
\affiliation{Physics department, Ecole Normale Supérieure, Université PSL, 24 rue Lhomond 75005 Paris France}
\author{Federico Centrone}
\affiliation{ICFO - Institut de Ciencies Fotoniques, The Barcelona Institute of Science and Technology, Castelldefels (Barcelona) 08860, Spain}
\author{Antonio Acín}
\affiliation{ICFO - Institut de Ciencies Fotoniques, The Barcelona Institute of Science and Technology, Castelldefels (Barcelona) 08860, Spain}

\date{\today}

\begin{abstract}
We discuss exchange scenario's thermodynamic uncertainty relations for the work done on a two-qubit entangled nonequilibrium steady state obtained by coupling the two qubits and putting each of them in weak contact with a thermal bath. 
In this way we investigate the use of entangled nonequilibrium steady states as end-points of thermodynamic cycles. 
In this framework, we prove analytically that for a paradigmatic unitary it is possible to construct an exchange scenario’s thermodynamic uncertainty relation. However, despite holding in many cases, we also show that such relation ceases to be valid when considering other suitable unitary quenches.
Furthermore, this paradigmatic example allows us to shed light on the role of the entanglement between the two qubits for precise work absorption. By considering the projection of the entangled steady state onto the set of separable states, we provide examples where such projection implies an increase of the relative uncertainty, showing the usefulness of entanglement.

\end{abstract}

\maketitle


\section{\label{sec:intro}
Introduction}
    In the last decades, owing to the continuous improvement in the miniaturization and control of devices down to the nanoscale, increasing efforts are devoted to understand small thermal engines, and if and how the laws of thermodynamics still apply in the microscopic world.
    In this scenario quantum mechanics naturally dictates the rules of the game.
    Quantum thermodynamics is becoming then a fundamental branch of modern physics \cite{anders2016reviewqtd, campisi2011rmp, alicki2018introduction} and  technology
    \cite{auffeves2022initiative}.
        
    When the size of the system  approaches the micro-nanoscale, the fluctuations  of thermodynamic variables become comparable or even larger than the expectation values, marking a relevant difference with macroscopic statistical mechanics. Finding bounds for the uncertainty means understanding the profound origin of the fluctuations and, more practically, helps monitoring the precision of a thermodynamic process.
    This interest led in the last years to the discovery of thermodynamic uncertainty relations (TURs) which provide a lower bound for the uncertainty of thermodynamic variables.
    It happens that the higher the precision one wants to achieve, the higher the dissipation cost.
    The first TURs were introduced by Seifert et al. during the last decade  \cite{baratoseifert2015tur, seifert2016pre, seifert2017pre, seifert2018steady}, however their generality and fundamental origin is still debated \cite{horowitz2020thermUR}.

    In Ref.\,\cite{landi2019tur} it was shown that specific TURs can be derived from the {\it exchange fluctuation theorem}  \cite{jarzynski2004exchange}.
    The  situation there considered concerns a thermodynamic cycle in which two (or more) energetically decoupled bodies (e.g. qubits) start each in an own thermal state achieved by weak interaction with a thermal bath.
    The initial state is then the tensor product of the two single-qubit equilibrium steady states (ESS). Once in this ESS the system has associated zero average entropy production \cite{avgEP} since no heat is exchanged with the thermal baths and the Von Neuman entropy is constant.
    Then an interaction Hamiltonian is turned on and then off, implying energy exchange between the two bodies. Finally, the cycle is closed by letting the two bodies thermalize again with their respective baths, implying irreversible entropy production.
    In this framework a function of the average irreversible entropy production produced in the relaxation step (i.e. before the ESS is reached) provides a lower bound for the relative error of thermodynamic \textit{charges}, e.g. work or exchanged heats. 
    Specifically, the TUR in the exchange scenario of Ref.\,\cite{landi2019tur} for the work $W$ is an inequality of the following form
\begin{equation}
\label{TUR-generic}
    \frac{\langle W^2\rangle-\langle W\rangle^2}{\langle W\rangle^2} \geq {\cal F}(\Sigma),
    \end{equation}
    where the function ${\cal F}(\Sigma)$ in the rhs is generally a monotonically decreasing positive function of the irreversible entropy production associated to the cycle [see Eq.~\eqref{bounds} for explicit forms of ${\cal F}$].\\

    With the aim of probing nonequilibrium and purely quantum features, we consider here a modified setup.
    The two bodies are energetically coupled, making the initial configuration a nonequilibrium steady state (NESS). A constant heat current flows from the hot bath to the cold bath mediated by the quantum system posed in the middle.
    At variance with the ESS configuration of \cite{landi2019tur}, (i)~the NESS implies a non-zero (constant) entropy production rate; 
    (ii)~not being under the hypotheses of the exchange fluctuation theorem, in principle there is no guarantee that a TUR will appear;
    (iii)~the initial NESS configuration may contain quantum correlations whose role deserves attention, hopefully driving an improvement in precision. 
    Indeed, it is a know fact that entanglement is useful for precision tasks in quantum metrology \cite{cappellaro2005entanglement,toth2012multipartite} hence intuitively it may be advantageous for the precision of the thermodynamic cycle.

    It should be noticed that other TURs for NESS were already studied in literature regarding the fluctuations of thermodynamic currents in the NESS, both in classical and quantum frameworks \cite{Horowitz2017rapidcomm, 2018-Coherence-qdots-ness-tur,guarnieri2019ness}. 
    Notice that the situation we consider is different: we consider a cycle in which the extreme points are described by NESS and a unitary sending the system out of the NESS. In this regard, our construction is closer to the scheme presented in \cite{landi2019tur, sacchi2021qudits} than to studies regarding current fluctuations in the NESS \cite{Horowitz2017rapidcomm, 2018-Coherence-qdots-ness-tur,guarnieri2019ness}.

    The main findings of our work are the following.
    In our framework, we first prove analytically that for a paradigmatic unitary an exchange scenario’s TUR still applies. However, despite holding in many cases, we also find that violations of such TUR can be achieved via other suitable unitary quenches.
Furthermore, we investigate the effect of the entanglement between the two qubits on the precision of work absorption. By considering the Euclidean projection of the NESS onto the set of separable states, we provide examples where such projection implies an increase of the relative uncertainty. This argument witnesses the usefulness of the presence of entanglement in the NESS.
    
    The article is organized as follows. In Sec.\,\ref{sec:model} we introduce our scheme based on the two-point-measurement setup and on the NESS as extreme points of the cycle. 
    We focus on the paradigmatic example of the two-qubit entangled NESS.
    In Sec.\,\ref{sec:analysis}, we derive a TUR for the work treating a particular unitary which, swapping the two entangled eigenstates of the Hamiltonian, deserves attention in our analysis.
    However, we show that in general the setup under consideration is not limited by the exchange scenario TURs.
    We finally discuss the role of entanglement in the framework.
    Our conclusions together with possible future extensions are presented in Sec.\,\ref{sec:conc}.


\section{\label{sec:model}
Preliminaries and NESS-based stroke absorber}
Here we introduce our original protocol that is based on a previously studied \cite{Hofer_2017, Khandelwal_2020} two-qubit nonequilibrium model.
To facilitate the reading, we report here some essential elements.
\subsection{Two-qubit model}
\begin{figure}
    \centering
    \includegraphics[width=1\columnwidth]{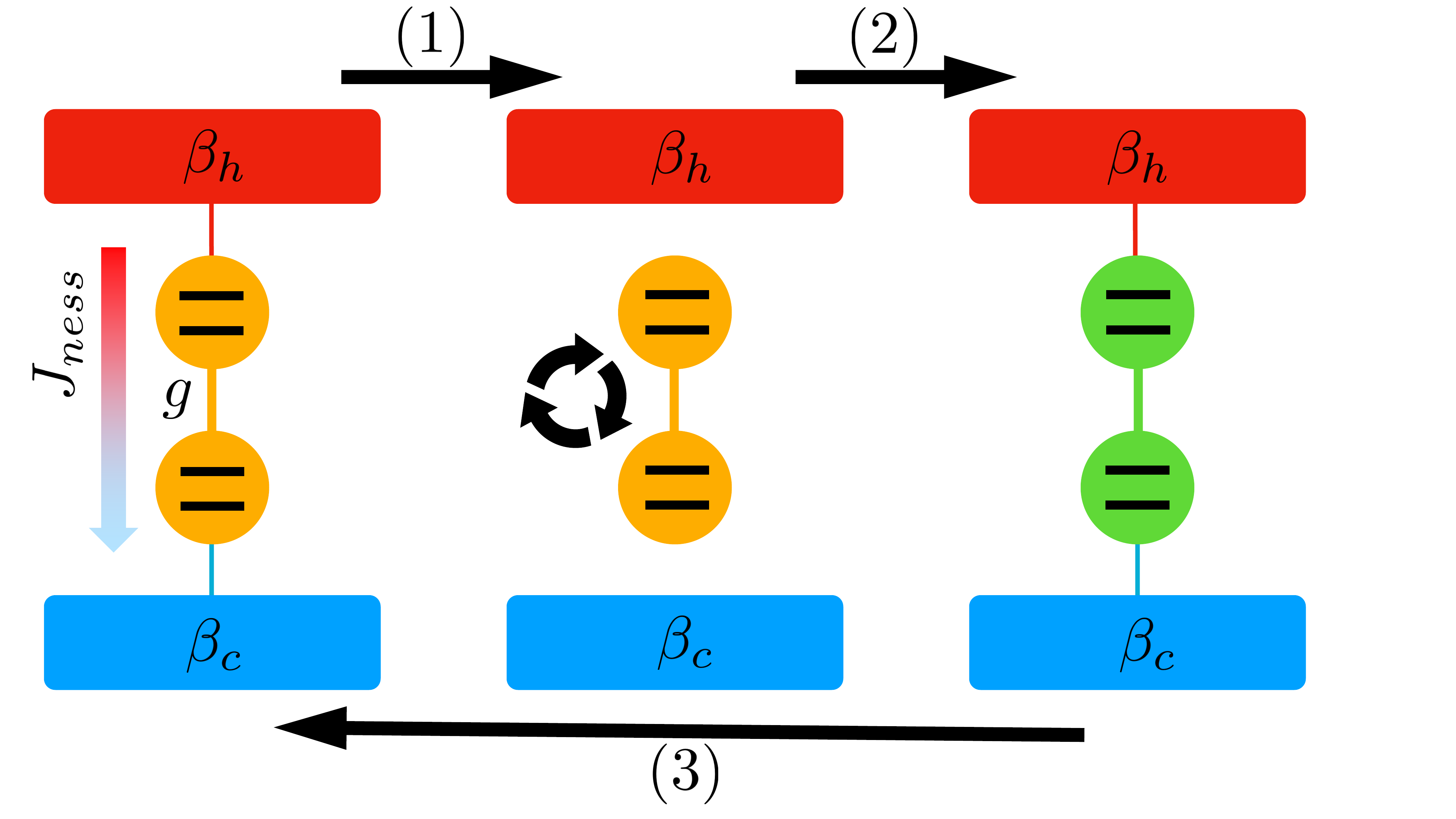}
    \caption{Schematic representation of the NESS-based stroke absorber. The system is composed of two coupled qubits (coupling constant $g$). At variance with conventional protocols the starting state of the system in the cycle is a NESS (left) which describes a constant heat current $J_{\text{ness}}$ from the hot bath (temperature $1/\beta_h$) to the cold bath (temperature $1/\beta_c$)  mediated by the quantum system.   A fast unitary is performed on the system (middle) which sends the state out of the NESS (right). This  allows to do work on the system. The system is finally left to reach again the initial NESS closing the cycle.}
    \label{fig:schema}
\end{figure}

The setup we study is schematically depicted in Fig.\,\ref{fig:schema}.
We consider a system $S$ composed of two qubits. 
The Hamiltonian of $S$ is 
\begin{equation}
H=H_0+H_{int} 
\,
,
\end{equation}
with
\begin{align}
	{H_0}&=\Omega_c \ket{1}\bra{1}\otimes\mathbb{I}_2 + \Omega_h \mathbb{I}_2 \otimes \ket{1}\bra{1}
	\end{align}
containing the local Hamiltonians of each subsystem and a coupling flip-flop term 
\begin{equation}
	{H}_{int}=g(\ket{01}\bra{10} + \ket{10}\bra{01})~.
\end{equation}
We consider for simplicity the resonant case $\Omega_c=\Omega_h\equiv\Omega$ that makes the eigenvalues of $H$ to be
\begin{equation}
\label{avals-H}
    (0, \Omega- g, \Omega+g, 2\Omega)
\end{equation}
 with associated eigenvectors
\begin{equation}
\ket{\varphi_0}=\ket{00}\,,\,
\ket{\varphi_{\Omega\pm g}}=\frac{\ket{01}\pm\ket{10}}{\sqrt{2}}\,,\,
\ket{\varphi_{2\Omega}}=\ket{11}\,.
\label{eigenvectors}
\end{equation}
\subsection{Dynamics}
%
We put each qubit in {weak} coupling with a own {Markovian} thermal bath. We name the two baths $B_c$ and $B_h$ with corresponding temperatures $1/\beta_c$ (colder) and $1/\beta_h$ (hotter), respectively. 
We also assume the thermal baths to be bosonic and interacting with the system through an excitation conserving interaction Hamiltonian, that couples the excitation of a qubit with the loss of a photon from its own bath and vice-versa. 
In this framework, the dynamics of $S$ is described by the  Gorini–Kossakowski–Sudarshan–Lindblad master equation~\cite{gorini1976completely,lindblad1976generators}.
Furthermore, we consider Ohmic spectral density and, more importantly, strong internal coupling with the aim of generating entanglement.
In this regime the global master equation yields an accurate description of the system state \cite{Hofer_2017, adesso2017locvsglob} and ensures the entropy production rate to be always non-negative \cite{ALICKI1976249, spohn1978semigroup}.
The master equation reads
\begin{multline}
\label{master-equation}
	\dot{\rho}(t) = -i\left[ {H}, {\rho}(t)\right] + \\ \sum_{\alpha=c,h} \sum_{\epsilon=\Omega\pm g}\,\,  
  \Gamma_{\alpha,\epsilon} \mathcal{D}[{L}_{\alpha,\epsilon}^\dagger]{\rho}(t) + \overline{\Gamma}_{\alpha,\epsilon} \mathcal{D}[{L}_{\alpha,\epsilon}]{\rho}(t)\,,  
	\end{multline}
where $ \mathcal{D}$ denotes the dissipator superoperator defined as
\begin{equation}
    \mathcal{D}[{L}]\rho =  L \rho  L^\dagger -\frac{1}{2} \{ L^\dagger  L,\rho\}
\end{equation}
and the transition rates are expressed by
\begin{equation}
    \Gamma_{\alpha,\epsilon} = \frac{\nu_\alpha \epsilon}{e^{\beta_\alpha \epsilon}-1}\,,\, \overline{\Gamma}_{\alpha,\epsilon} = \nu_\alpha \epsilon \frac{e^{\beta_\alpha \epsilon}}{e^{\beta_\alpha \epsilon}-1}\,,
\end{equation}
with $\nu_\alpha$ being the coupling strength to each thermal bath.
According to this setup, only transitions with energies $\Omega\pm g$ can be induced by the system-bath coupling considered here \cite{Hofer_2017}, with 
corresponding jump operators being
\begin{align}
	{L}_{c,\Omega-g}&=\frac{1}{\sqrt{2}}\ket{\varphi_{\Omega+g}}\bra{\varphi_{2\Omega}}-\frac{1}{\sqrt{2}}\ket{\varphi_0}\bra{\varphi_{\Omega-g}}\\
	{L}_{c,\Omega+g}&=\frac{1}{\sqrt{2}}\ket{\varphi_{\Omega-g}}\bra{\varphi_{2\Omega}}+\frac{1}{\sqrt{2}}\ket{\varphi_0}\bra{\varphi_{\Omega+g}}\\
	{L}_{h,\Omega-g}&=\frac{1}{\sqrt{2}}\ket{\varphi_{\Omega+g}}\bra{\varphi_{2\Omega}}+\frac{1}{\sqrt{2}}\ket{\varphi_0}\bra{\varphi_{\Omega-g}}\\
	{L}_{h,\Omega+g}&=-\frac{1}{\sqrt{2}}\ket{\varphi_{\Omega-g}}\bra{\varphi_{2\Omega}}+\frac{1}{\sqrt{2}}\ket{\varphi_0}\bra{\varphi_{\Omega+g}}\,.
	\end{align}
The NESS, i.e. the stationary state of Eq.~\eqref{master-equation}, 
is diagonal in the Hamiltonian's eigenbasis \eqref{eigenvectors},
assuming the form

\begin{eqnarray}
\label{ness}
    {\rho}_{\text{ness}} =\rho_0 \ket{\varphi_0}\bra{\varphi_0}+\\
    \rho_- \ket{\varphi_{\Omega-g}}\bra{\varphi_{\Omega-g}}+\nonumber
    \rho_+\ket{\varphi_{\Omega+g}}\bra{\varphi_{\Omega+g}}+\nonumber\\
    \rho_{2\Omega} \ket{\varphi_{2 \Omega}}\bra{\varphi_{2 \Omega}}\nonumber\,.
\end{eqnarray}
The expressions of the coefficients ${\rho}_0, \rho_-, \rho_+, \rho_{2 \Omega}$ are reported in Appendix~\ref{app:steadystate} (see also  \cite{Khandelwal_2020}) for brevity.
Here we simply remark that the NESS is a passive state, meaning that no work can be extracted from it (see Appendix~\ref{app:passivity} and  \cite{Khandelwal_2020} for further details), as a consequence of the fact that the populations are decreasing in the energy eigenvectors, namely ${\rho}_0\geq \rho_- \geq  \rho_+ \geq  \rho_{2 \Omega}$. 
\subsection{Thermodynamic cycle}
For long enough time, the interaction with the baths sends $S$ in a NESS, with density matrix $\rho_{\text{ness}}$, which describes a constant heat current $J_{\text{ness}}$ from the hot bath to the cold bath. 
The NESS is the initial state of the protocol we present here.
In order to do work $W$ on the system,
at time $t=-\delta t$ we apply a fast unitary operation $U$  which sends the system out of the NESS.
With \textit{fast} we mean that the external field is turned on and off on timescales $\delta t$ much smaller than the relaxation timescales.
Alternatively, one may think to decouple the system from the thermal baths during the work step which now can occur on arbitrary timescales.
In both the cases, the resulting state at $t=0$ is then 
\begin{equation}
    \rho(0)=U \rho_{\text{ness}} U^\dag\,.
\end{equation}
In the following we shall use the convention that work (heat) is positive when it is done on (absorbed by) the system.
The work probability distribution is constructed through the two-point-measurement scheme \cite{tasaki2000jarzynski, kurchan2000quantum, Mukamel2000prlqjarz, hanggi2007WnotObs, anders2016reviewqtd}
and reads
\begin{eqnarray}
\label{pdfw}
&& P(W)=\\
&&
\sum_{n,m} \bra{\varphi_{\epsilon_n}}\rho_{\text{ness}}\ket{\varphi_{\epsilon_n}}
\vert\bra{\varphi_{\epsilon_m}}U\ket{\varphi_{\epsilon_n}}\vert^2\delta(\epsilon_m-\epsilon_n-W)\,, \nonumber
\end{eqnarray}
where the first energy measurement is performed on the NESS (time $t=-\delta t$) and the second just after the application of the unitary (time $t=0$). 
Notice that the Hamiltonian of the system is the same (i.e., $H$) at the two measurement points.
Eq.\,\eqref{pdfw} allows one to calculate the statistical moments of $W$.
Defining $\langle{f(W)}\rangle:=\int_{-\infty}^{+\infty} dW P(W) f(W)$, the average work is $\langle{W}\rangle$, while the work variance is  
$Var(W)=\langle{W^2}\rangle-\langle W\rangle ^2$.

The cycle closes by allowing the system to relax with the two baths, finally reaching the initial NESS (symbolically, time $t \rightarrow \infty$, meaning at times much greater than the relaxation time imposed by \eqref{master-equation}), namely we have 
\begin{equation}
    \rho(\infty)=\rho_{\text{ness}}
\end{equation}
as end-point of the cycle.

\subsection{Entropies}
In this subsection we introduce entropic quantities that will be useful for subsequent discussions.
In particular, considering our NESS-based cycle, we are interested in identifying sensible candidates for the entropy $\Sigma$ entering in the rhs of Ineq.\,\eqref{TUR-generic}.

To start with, the Von Neuman entropy  of $S$ is the state function  
\begin{equation}
    S(t)=-{\rm Tr}[\rho(t) \ln \rho(t) ]
\end{equation}
and
its temporal change 
$\Delta S(t)=S(t)-S(0)$,
has reversible and irreversible contributions,
\begin{equation}
\label{balance-entropies}
    \Delta S(t)=\Delta_e S(t)+\Delta_i S(t)\,.
\end{equation}
The term 
\begin{equation}
\label{entropy-e}
\Delta_e S= \beta_h \langle Q_h \rangle+\beta_c \langle Q_c \rangle
\end{equation}
is the reversible contribution due to the heat exchanges with the thermal baths \cite{Esposito_2010, barra2015thermodynamic}
and $\Delta_i S$ is the irreversible contribution called entropy production.
We also define the time derivative of \eqref{entropy-e}, i.e. the entropy flow from the environment to the system, as
\begin{equation}
\label{entropyflow}
\dot{S}_e(t)= \beta_h J_h(t)+\beta_c J_c(t)    
\end{equation}
where  
\begin{align}
     &J_\alpha(t)=\langle \dot{Q}_\alpha \rangle(t)=\\
\nonumber   & =\text{Tr}\{H  \sum_{\epsilon=\Omega\pm g}\,   \Gamma_{\alpha,\epsilon} \mathcal{D}[{L}_{\alpha,\epsilon}^\dagger]{\rho}(t) + \overline{\Gamma}_{\alpha,\epsilon} \mathcal{D}[{L}_{\alpha,\epsilon}]{\rho}(t) \}
\end{align}
is the heat current associated to bath $\alpha$. 
 From Eq.~\eqref{balance-entropies} the entropy production rate then reads 
\begin{equation}
\label{balance-entropies-rates}
    \dot{S}_i(t)=\dot{S}(t)-\dot{S}_e(t)~.
\end{equation}

Regarding the protocol described in the previous subsection,
after a cycle, by definition, we have a total null entropy change, $\Delta S\vert_{cycle}:=S(\infty)-S(0)=0$, implying $\Delta S_i\vert_{cycle}=-\Delta S_e\vert_{cycle}$.
Notice that, for arbitrarily large cycle's time, $\Delta S_i\vert_{cycle}=\Delta_i S(t \rightarrow
\infty)$ diverges because its asymptotic rate is non-zero.
On the contrary, the entropy production excess with respect to the NESS value, i.e. the dissipation cost needed for applying the unitary $U$, can remain finite.
In formulas, making explicit the dependence on the initial state, we define
\begin{equation}
\Delta S_{cost}(t):=
\Delta_i S(t; U\rho_{\text{ness}}U^\dag)-\Delta_i S(t; \rho_{\text{ness}})\,,
\label{Scost}
\end{equation}
where we subtract to the actual term $\Delta_i S(t; U\rho_{\text{ness}}U^\dag):=\Delta_i S(t)$, the constantly growing term we would have 
obtained without applying the unitary $U$.
In other words, $\Delta_i S(t; \rho_{\text{ness}})$
is treated as an offset, where the state remains in the NESS for all $t$. 
If we consider the associated rate
\begin{equation}
\dot{S}_{cost}(t):=
 \dot{S}_i(t; U\rho_{\text{ness}}U^\dag)- \dot{S}_i(t; \rho_{\text{ness}})\,,
 \label{Scost-rates}
\end{equation}
it tends to nullify as the state tends to the NESS for $t\rightarrow \infty$.

Consequently, for our NESS-based cycle, while $\Delta_i S\vert_{cycle}$ cannot enter in the rhs of \eqref{TUR-generic} (being diverging yields for the rhs the trivial value zero),
the steady state value of \eqref{Scost}
\begin{equation}
\label{Sigma-cost}
    \Sigma_{cost}:=\Delta S_{cost}(\infty)
\end{equation}
is instead a possible sensible (finite) candidate.
Introducing the quantum relative entropy
\begin{equation}
    \text{S}(\rho||\sigma)=\text{Tr}[ \rho (\ln \rho-\ln \sigma) ]
\end{equation}
another well defined candidate is
\begin{equation}
\label{Srel}
   \Sigma_{rel}:=\text{S} (U\rho_{\text{ness}} U^\dagger||\rho_{\text{ness}})
\end{equation}
which geometrically is a quantifier of the distance between the density operators $U\rho_{\text{ness}}U^\dagger$
and
$
\rho_{\text{ness}}$ and physically corresponds to the nonadiabatic contribution to the entropy production in the relaxation process \cite{spohn1978entropy, PhysRevE.92.032129, PhysRevX.8.031037}.
The fact that both \eqref{Sigma-cost} and \eqref{Srel} could represent eligible entropies entering in \eqref{TUR-generic} is encouraged by the fact that, in a conventional stroke-based engine with ESS, they both
reduce to the same quantity, the average entropy production, and take the role of $\Sigma$ entering in the rhs of \eqref{TUR-generic}  \cite{landi2019tur}. 
\begin{figure}
\centering
\begin{overpic}[width=.9\linewidth]{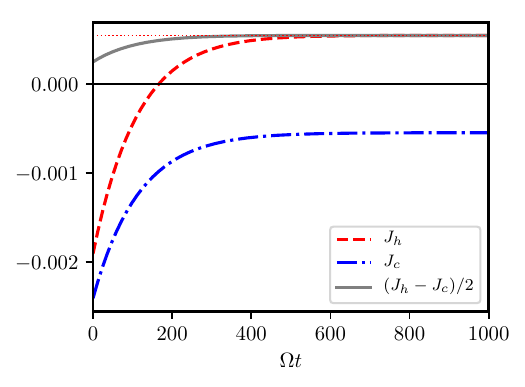}
\put(20,75){(a)}
\end{overpic}\\
\vspace{.5cm}
\begin{overpic}[width=.9\linewidth]{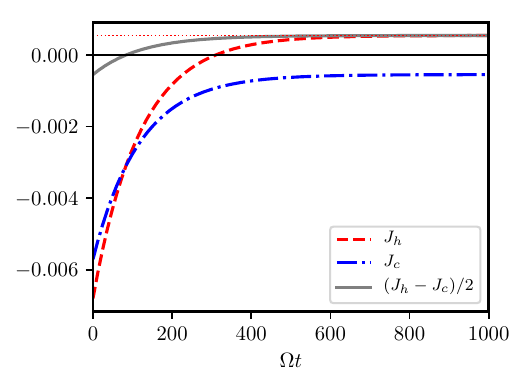}
\put(20,75){(b)}
\end{overpic}
    \caption{(a) Having used the unitary operation defined in \eqref{swapent} to perform work on the system, we plot the
    hot and cold heat currents during relaxation to the steady state for $\beta_c = 3, \beta_h = 1, \nu_c=\nu_h=0.004$ with energy levels $\Omega=1$ and $g=0.75$. During the transient regime, $(J_h-J_c)/2$ is smaller than in the NESS, meaning a net slower heat transfer between the two baths.
    (b) Same but with the maximum work unitary \eqref{maxworkunitary}. Notice the negative value of $(J_h-J_c)/2$ at short timescales, meaning more heat given to the hot bath than to the cold bath at those timescales.
    }
    \label{fig:J}
\end{figure}

\begin{figure}
\centering
\begin{overpic}[width=.94\linewidth]{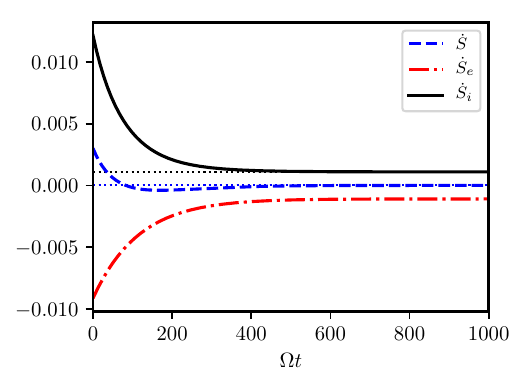}
\put(30,60){}
\end{overpic}
    \caption{Using the unitary in \eqref{swapent}, relaxation of the entropy rates to the NESS values. 
    $\dot{S}(t)$ is the derivative of the Von Neuman entropy of the state and the entropy flow $\dot{S}_e(t)$ is computed from the heat currents as in \eqref{entropyflow}. 
    The entropy production rate $\dot{S}_i(t)$ is given by subtracting the latter to the first term, as in  \eqref{balance-entropies-rates}.
    We used the parameters $\Omega=1$, $\beta_c = 3, \beta_h = 1, \nu_c=\nu_h=0.004$ and $g=0.75$.
    The NESS is characterized by a positive entropy production rate (black dotted). 
    The unitary implies an entropy production cost during the time transient with respect to the NESS value, geometrically the area between black full and black dotted lines.
    \label{fig:entropy-rates}
    }
\end{figure}

\section{\label{sec:analysis}
Analysis}
\subsection{Precision using selected quenches}
\begin{figure}
    \centering
    \begin{overpic}[width=.86\linewidth]{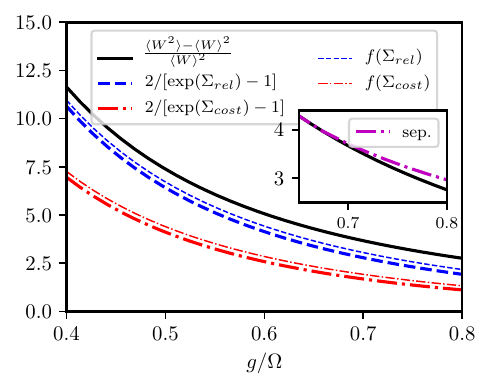}
\put(20,80){(a)}
\end{overpic}\\
\vspace{.5cm}
\begin{overpic}[width=.86\linewidth]{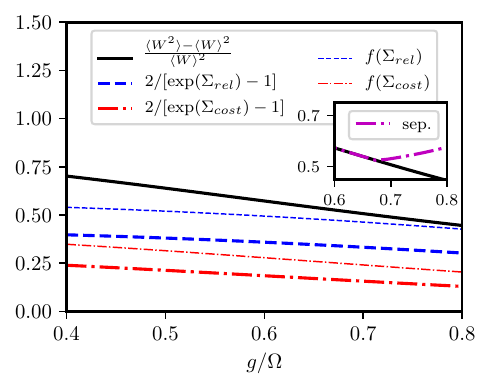}
\put(20,80){(b)}
\end{overpic}
\caption{(a) Using the unitary
\eqref{swapent} and setting
$\Omega=1$, $\nu_c=\nu_h=0.004$, $\beta_c=3$, $\beta_h=1$
we plot the lhs and rhs of \eqref{TUR-generic} as function of the internal coupling $g$, considering the four different forms for the rhs as detailed in \eqref{bounds}.
(b) Same as in (a), but using the maximum work unitary \eqref{maxworkunitary}. This leads to lower relative error.
In both (a) and (b), under the selected settings no violation of the exchange scenario's TURs are found and the precision increases for increasing internal coupling $g$.
The insets show instead the behavior of the relative error substituting the NESS with the closest separable state in the Frobenius norm.
In the latter case the relative error increases, an argument in favour of the usefulness of entanglement.
\label{fig:TURs}}
\end{figure}

\subsubsection{Swap in the entangled basis}
In the exchange scenario of Ref.\,\cite{landi2019tur}
the relative uncertainty of work is lower bounded by a function of the average entropy production. 
To start with, being interested in possible beneficial interplay between correlations and precision, we set the unitary to be
\begin{eqnarray}
\label{swapent}
    U:=\ket{\varphi_0}\bra{\varphi_0}+\\
    \ket{\varphi_{\Omega+g}}\bra{\varphi_{\Omega-g}}+\nonumber
    \ket{\varphi_{\Omega-g}}\bra{\varphi_{\Omega+g}}+\nonumber\\
    \ket{\varphi_{2 \Omega}}\bra{\varphi_{2 \Omega}}\nonumber\,.
\end{eqnarray}
It acts non-trivially only on the entangled part of the energy basis, swapping the two entangled eigenstates. Furthermore, notice that it is non-local and can in principle build or destroy entanglement in a generic bipartite quantum state. However, it can be shown that for any state written in the form of our NESS in Eq.~\eqref{ness} the quench \eqref{swapent} does not modify the amount of entanglement.

\textit{Dynamics.---}To get insights on the relaxation process, solving the master equation \eqref{master-equation} we plot  
in Fig.~\ref{fig:J}(a) the currents
$J_h(t)$ and $J_c(t)$
as function of time, together with their rescaled difference (the chosen parameters are indicated in the figure's caption) \cite{johansson2012qutip}.
The negative current values express the fact that in the relaxation process at short times the system provides heat to both the baths.  
In Fig.~\ref{fig:entropy-rates} we plot instead
the corresponding relaxations to the NESS values of the reversible and irreversible contributions to the system's entropy rate entering in the expression \eqref{balance-entropies-rates}.
The NESS is characterized by a positive entropy production rate $\dot{S}_i(\infty)$, black dotted line.  
The application of the unitary has an entropy production cost $\Sigma_{cost}$ that is geometrically the area delimited by $\dot{S}_i(t)$ (black continuous line) and $\dot{S}_i(\infty)$.

\textit{Existence of a TUR.---}For the particular unitary \eqref{swapent}, a bound for the work relative uncertainty still applies as we shall discuss now.
The relative uncertainty of the work is lower bounded 
by the function introduced in
\cite{Merhav_2010,Proesmans_2017,hasegawa2019fe}, i.e. $f_0(x):={2}/({e^{x}-1})$,
where $x$ must be set equal to the quantum relative entropy between the NESS evolved with the unitary $U$ and the NESS itself, namely
\begin{equation} \label{TUR}
    \frac{\langle W^2\rangle-\langle W\rangle^2}{\langle W\rangle^2} \geq \frac{2}{e^{\Sigma_{rel}}-1}~,
        \end{equation}
with $\Sigma_{rel}$ defined in \eqref{Srel}.
The proof of Ineq.~\eqref{TUR} is reported in Appendix~\ref{app:prooftur} and both the lhs and rhs are plotted in Fig.~\ref{fig:TURs}(a) as function of the the internal coupling $g$.

More generally, we plot in Fig.~\ref{fig:TURs}(a) the rhs of the TUR  corresponding to the several bounds established in the exchange scenario. Such bounds are defined by the functional form of $\mathcal{F}(x)$ appearing in the rhs of Ineq.\,\eqref{TUR-generic} and by the argument $x$,
namely $\mathcal{F}\in \{f_0, f\}$, with
\begin{eqnarray}
&&f_0(x):={2}/({e^{x}-1})\,,
\label{bounds}\\
\nonumber
&&f(x):=1/\sinh^2[y \, {\rm s.t.}  \,(y \tanh(y)=x/2)],\\
&& x\in \{\Sigma_{rel}, \Sigma_{cost}\}\,.\nonumber
\end{eqnarray}
The function $f(x)$ was introduced in Ref.\,\cite{landi2019tur}, representing the tightest saturable bound in the exchange scenario, implying automatically $f(x)\geq f_0(x)$.
We numerically observe that in our nonequilibrium framework the relative entropy is always lower than the entropy production cost, despite the two quantities coincide in the exchange scenario \cite{landi2019tur}. 
We notice that all the bounds are respected in
Fig.~\ref{fig:TURs}(a).
\subsubsection{Maximum work unitary}
With the aim of considering the optimal process,
we can turn our attention to the unitary corresponding to the maximum  work doable on the system,
\begin{eqnarray}
\label{maxworkunitary}
    U\equiv\ket{\varphi_0}\bra{\varphi_{2 \Omega}}+\\
    \ket{\varphi_{\Omega+g}}\bra{\varphi_{\Omega-g}}+\nonumber
    \ket{\varphi_{\Omega-g}}\bra{\varphi_{\Omega+g}}+\nonumber\\
    \ket{\varphi_{2 \Omega}}\bra{\varphi_{0}}\nonumber\,.
\end{eqnarray}
Interestingly, from Fig.~\ref{fig:J}(b), while providing energy to both baths during the time transient, the system supplies more energy to the hot bath during a small time transient (negative values of the gray continuous curve). 
As shown in Fig.~\ref{fig:TURs}(b) and comparing it with Fig.~\ref{fig:TURs}(a), 
the maximum work unitary also allows for better precision while  still not violating the exchange scenario bounds for the selected parameters in the plot.
\subsubsection{Violation of the exchange scenario's TURs}

The violation of TURs in nonequilibrium classical and quantum scenarios has recently been studied in different settings \cite{agarwalla2018assessing,cangemi2020violation,paneru2020reaching}. A similar analysis, however, was never performed in nonequilibrium settings for the exchange scenario's TURs. Indeed, in our {\it nonequilibrium exchange scenario} the hypotheses of the exchange fluctuation theorem do not hold and thus the existence of such TURs is not guaranteed.  
As a matter of fact, we demonstrate that all the bounds in \eqref{bounds} can  indeed be violated with a suitable choice of the unitary operation. This is shown in Fig.~\ref{fig:TUR-breaking}. The form of the selected unitary is written explicitly in Appendix~\ref{app:randU}.

\begin{figure}
    \centering
\begin{overpic}[width=.85\linewidth]{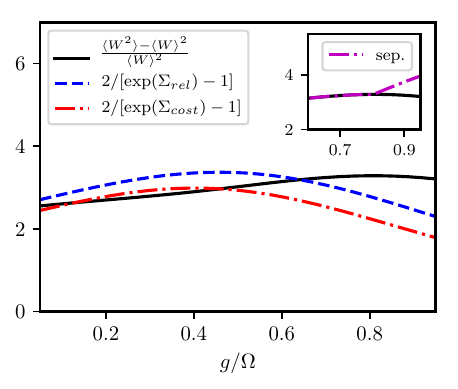}
\put(20,85){}
\end{overpic}
\caption{
Setting $\Omega=1$, $\nu_c=0.002$, $\nu_h=0.008$, $\beta_c=3$, $\beta_h=1$
we plot the lhs and rhs of \eqref{TUR-generic}, using the first expression $f_0(x)$ in \eqref{bounds} for both $x=\Sigma_{cost}$ and $x=\Sigma_{rel}$ (see legend) and a unitary that allows one to violate both the bounds, reported in Eq.~\eqref{Uviol}.
Here the relative error does not decrease monotonically with $g$.
The inset shows instead the behavior of the relative error substituting the NESS with the closest separable state in the Frobenius norm.
In the latter case the relative error increases, an argument in favour of the usefulness of entanglement, similarly to Fig.\,\ref{fig:TURs}.
\label{fig:TUR-breaking}
}
\end{figure}
This means that, in general, in the setup under consideration the precision is not limited by the exchange scenario TUR bounds, namely there exist parameter settings for which 
$({\langle W^2\rangle-\langle W\rangle^2})/{\langle W\rangle^2} < {\cal F}(\Sigma)
$,
for all $\Sigma$ and ${\cal F}(\Sigma)$ defined in \eqref{bounds}. 
In particular, we emphasize that a saving in entropy production cost as defined in Eq.\,\eqref{Sigma-cost} does not limit the precision, hence marking a difference with respect to the exchange scenario discussed in \cite{landi2019tur}.

\textit{Rare violations.---} In order to obtain more insights on the validity of the TURs in different nonequilibrium exchange scenarios, we perform computations for a large number of unitaries drawn from a Haar measure.
\begin{figure}
    \centering
    \includegraphics[width=1\columnwidth]{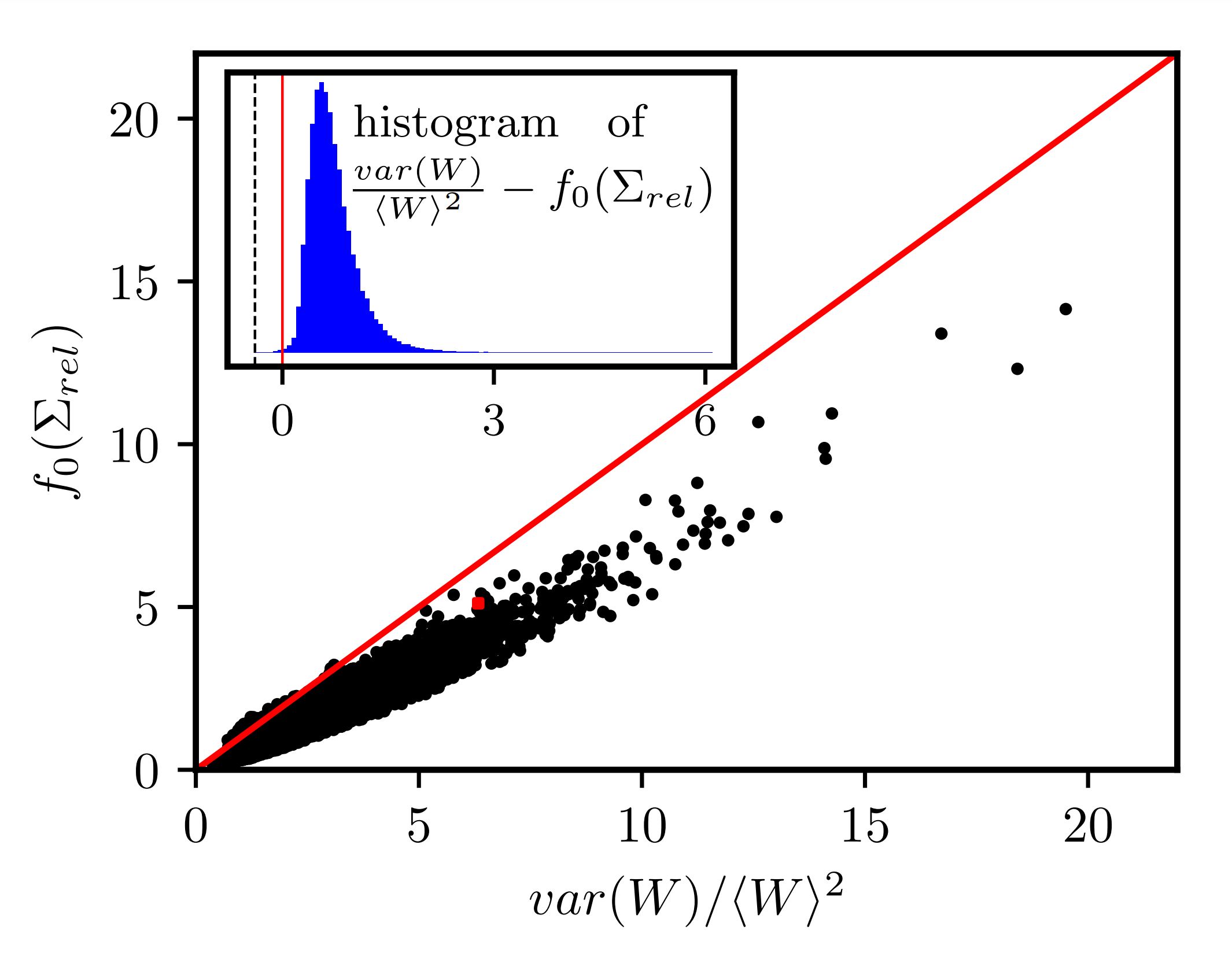}
    \caption{TUR evaluation for 100000 random unitaries drawn from a  Haar measure for $\beta_h=1, \beta_c=3, \nu_h=0.004, \nu_c=0.012, \Omega=1, g=0.5$. The work relative error squared (abscissa) corresponds to the lhs of the TUR, while for the rhs we take $f_0(\Sigma_{rel})$ (ordinate). The violations happen mostly in the regime of high thermodynamic precision, points above the red line. The red point indicates the unitary from Eq.~\eqref{swapent}. Inset: histogram of lhs-rhs. The bins below zero (vertical red line) quantify the small amount of violations observed, a fraction $\approx 0.0017$. The vertical dashed line corresponds to the maximum violation observed, corresponding to a value $\approx-0.39$.}
    \label{fig:Uviol}
\end{figure}
The analysis is reported in Fig.\,\ref{fig:Uviol}.
We numerically observe that violations are rare and manifest especially in the regime where both the lhs and rhs take small values (high precision). 
Also, the violations we found in all cases are not arbitrarily large, hence not excluding the possibility that a looser TUR bound might hold in our settings.
\subsection{Precision boost from entanglement}
The NESS obtained in this setup is in general non-separable \cite{Hofer_2017, adesso2017locvsglob, Khandelwal_2020} for large enough internal coupling $g$.
Specifically, the NESS \eqref{ness} is entangled iff \cite{Khandelwal_2020}
\begin{equation}
(\rho_--\rho_+)^2>4 \rho_0 \rho_{2 \Omega}\,.
\label{ent-cond}
\end{equation}
We remark that in our framework it is immediate to build entanglement criteria (equivalent to the above condition) that are based on thermodynamic quantities. 
For instance, for the specific case of the unitary  \eqref{swapent} we find the following necessary and sufficient condition for the presence of entanglement in the steady state:
\begin{equation}\label{thermo_ent-cond}
2\gamma>2(\xi-1)^2+\xi^2~,
\end{equation}
where  $\xi:=\langle W^2\rangle / (4g^2)=\rho_++\rho_-$, i.e. $\xi$ is proportional to the mean squared work, while $\gamma:=\text{Tr}(\rho^2_{\text{ness}})$ is the purity of the NESS.
Further developments are reported in Appendix~\ref{app:ent-criterio}.
The born of entanglement for large enough internal coupling $g$ is reported instead in the Appendix in Fig.~\ref{fig:NESS-properties} using the concurrence as quantifier.
This allows us to shed light on the interplay between quantum correlations and precision in thermodynamics.
Additionally, in Fig.~\ref{fig:NESS-properties} we plot, beyond the concurrence, the mutual information and the purity of the NESS as function of the internal coupling $g$.
Comparing it with Fig.~\ref{fig:TURs}, 
we observe that increasing correlations and the formation of entanglement accompany the reduction of the relative error and that this is not due to the fact that the state becomes purer, since the purity, in contrast, decreases. In addition, the selected unitaries \eqref{swapent} and \eqref{maxworkunitary} do not change the entanglement in the NESS, thus the advantage observed in the reduction of the relative work uncertainty may be due to the quantum correlations already present in the state.  

We provide the following argument in favor of the usefulness of entanglement. 
If we replace the entangled NESS with the closest separable state (i.e., the \enquote{classically constrained state}), we should observe some increase in the relative error to sustain the hypothesis that entanglement is useful in this scenario. 
Therefore, we consider the closest separable state $\bar{\rho}_{\text{sep}}$ in the Frobenius norm. To evaluate it we solve the following convex optimization problem,
\begin{equation}
\label{minimization-main}
\bar{\rho}_{\text{sep}}:=arg\left(
\begin{array}{rrclcl}
\displaystyle \min_{\rho_{\text{sep}}} & \multicolumn{3}{l}{{\lVert \rho_{\text{ness}}-\rho_{\text{sep}}   \rVert}_2}\\
\textrm{s.t.} 
& \rho_{\text{sep}} \geq 0,\,{\rm Tr}(\rho_{\text{sep}})=1\\
& \rho_{\text{sep}}^{PT}\geq 0 
\end{array}\qquad\quad\right)\,.
\end{equation}
The analytical way to find the solution $\bar{\rho}_{\text{sep}}$ of the above convex optimization problem was introduced in Ref.\,\cite{Verstraete2002} and is detailed in Appendix~\ref{Appendix-minimization}.
From the thermodynamic point of view, the result is that when replacing $\rho_{\text{ness}}$ with $\bar{\rho}_{\text{sep}}$,
the relative error of the work increases. This is shown in the insets of Fig.~\ref{fig:TURs} and Fig.~\ref{fig:TUR-breaking},
{where the black continuous lines refer to the actual behavior of the relative variance, while the dash-dotted magenta lines refer to the same quantity but obtained when using the closest separable state $\bar{\rho}_{\text{sep}}$ in the two-point-measurement scheme of equation \eqref{pdfw}.
Intuitively, the magenta lines represent the (worse) performance that would have been achieved if the state had been forced to stay inside the set of separable states. While not representing a definitive assessment on the importance of the entanglement in thermodynamic precision, our argument shows that, at least for paradigmatic settings, entanglement is a resource in the scheme we have discussed.
}
\\
\section{\label{sec:conc}
Conclusions}
Motivated by the growing interest in finding delineated regions of validity for the several TURs \cite{horowitz2020thermUR} and in understanding the role of purely quantum features in the context \cite{PhysRevLett128140602}, 
we studied the work relative uncertainty for a NESS-based stroke absorber.  
More specifically, we considered the paradigmatic NESS obtained by two coupled qubits, each in weak contact with a own thermal bath, with the two baths being at disparate temperatures. 
Our main findings are the following:
(i) Despite we show that an exchange scenario's TUR is valid when choosing specific unitary quenches  to describe the work absorption by $S$, in general if we are allowed to consider generic unitaries we numerically observe that the setup under consideration is not constrained by such TURs,
but violations are rare.
(ii) 
Investigating the effect of the entanglement between the two qubits on the work precision,
we find that entanglement is useful, at least in paradigmatic settings. 
Our statement is based on a comparison with the Euclidean projection of the NESS onto the set of separable states. We find that such projection implies an increase of the relative uncertainty, witnessing the usefulness of the presence of entanglement in the NESS.
Finally, it should be noted that
our study is restricted to the case where the two qubits have the same level spacing and to the Born-Markov-Secular approximation for the open system dynamics. 
Relaxations of one or more of such assumptions could represent natural developments.
\\

\begin{acknowledgments}
We thank Gonzalo Manzano, Vasco Cavina and Vittorio Giovannetti for fruitful discussions.
This work is supported by the Government of Spain (FIS2020-TRANQI and Severo Ochoa CEX2019-000910-S), Fundació Cellex, Fundació Mir-Puig, Generalitat de Catalunya (CERCA), the ERC AdG CERQUTE and the AXA Chair in Quantum Information Science.
\end{acknowledgments}

\begin{appendix}

\section{\label{app:steadystate}Global master equation's steady state}
\begin{figure}
    \centering
\begin{overpic}[width=.85\linewidth]{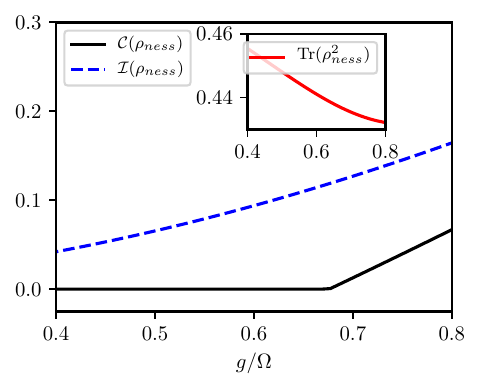}
\put(20,80){}
\end{overpic}
    \caption{
    Concurrence (black), mutual information (blue dashed) and purity (inset) of the NESS for $\Omega=1$, $\nu_c=\nu_h=0.004$ and $\beta_h=1,\beta_c=3$ as functions of the internal coupling $g$.
    Notice that for small values of $g$ the concurrence is 0. Remarkably, increasing correlations are accompanied by a reduction of the purity of the two-qubit state.
    }
    \label{fig:NESS-properties}
\end{figure}

{
The coefficients of the nonequilibrium steady state \eqref{ness} are \cite{Khandelwal_2020} 
\begin{eqnarray}
\label{ness-comps}
     \rho_0&=&a (\overline{\Gamma}_{c,\Omega-g}+\overline{\Gamma}_{h,\Omega-g})(\overline{\Gamma}_{c,\Omega+g}+\overline{\Gamma}_{h,\Omega+g}) \,,\\
    \rho_-&=&a (\Gamma_{c,\Omega-g}+\Gamma_{h,\Omega-g})(\overline{\Gamma}_{c,\Omega+g}+\overline{\Gamma}_{h,\Omega+g})\,,\nonumber\\
    \rho_+&=&a (\overline{\Gamma}_{c,\Omega-g}+\overline{\Gamma}_{h,\Omega-g})(\Gamma_{c,\Omega+g}+\Gamma_{h,\Omega+g})\,,\nonumber\\
    \rho_{2 \Omega}&=&a (\Gamma_{c,\Omega-g}+\Gamma_{h,\Omega-g})(\Gamma_{c,\Omega+g}+\Gamma_{h,\Omega+g})\,,\nonumber
    \end{eqnarray}
where the common proportionality  coefficient $a$ is the positive normalization quantity ensuring ${\rm Tr}({\rho}_{\text{ness}}) = 1$.
Purity and correlation properties of the NESS are described in Fig.\,\ref{fig:NESS-properties} for selected values of the system parameters.
}
\begin{figure}
\centering
\begin{overpic}[width=.85\columnwidth]{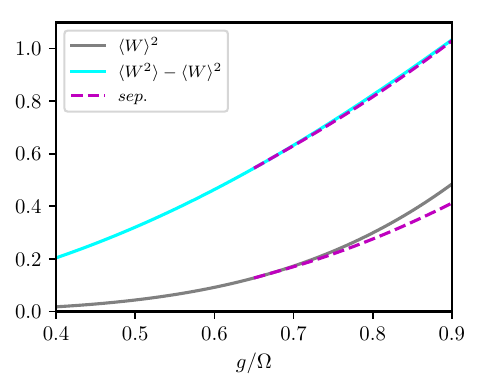}
\end{overpic}
\caption{
Squared average work and work variance as a function of $g$, for $\Omega=1$, $\beta_c = 3, \beta_h = 1, \nu_c=\nu_h=0.004$, considering the unitary \eqref{swapent}. 
Dashed lines correspond to the values obtained replacing the NESS with its closest separable state, see \eqref{minimization-main}: the average work decreases while the variance remains almost unchanged, implying an increase in the relative error.
}
\label{fig:workvariance}
\end{figure}
\section{Passivity}
\label{app:passivity}
The concept of ergotropy was introduced in \cite{Allahverdyan_2004} as the maximum extractable work by means of a driving which is turned on (start of work extraction) and then off (end of work extraction).
For the populations in \eqref{ness-comps},
it happens that
\begin{equation}
\label{populations-ordering}
    \rho_{2\Omega}\leq\rho_+\leq\rho_-\leq\rho_0\,.
\end{equation}
This implies that the steady state is passive.
\eqref{populations-ordering} can be verified by noticing that
\begin{align}
    \Gamma_{\alpha,\Omega+g}&\leq\Gamma_{\alpha,\Omega-g}\\
    \overline{\Gamma}_{\alpha,\Omega-g}&\leq\overline{\Gamma}_{\alpha,\Omega+g}\\
    \Gamma_{\alpha,\epsilon}&\leq\overline{\Gamma}_{\alpha,\epsilon}
\end{align}
where $\alpha=c,h$ and $\epsilon=\Omega\pm g$.
\section{\label{app:prooftur}Proof of the thermodynamic uncertainty relation \eqref{TUR}}
{
With the aim of proving \eqref{TUR}, we can start to evaluate the lhs. We get in terms of the NESS components in \eqref{ness} 
\begin{equation}
\begin{split}
    \langle W \rangle &= \text{Tr}\left[H(U\rho_{\text{ness}} U^\dagger - \rho_{\text{ness}})\right]\\
    &=2g(\rho_--\rho_+)
\end{split}
\end{equation}
and
\begin{equation}
    \langle W^2 \rangle =4g^2(\rho_++\rho_-)\,.
\end{equation}
The lhs hence reads
\begin{equation}
    \frac{\langle W^2\rangle-\langle W\rangle^2}{\langle W\rangle^2} = \frac{\rho_-+\rho_+}{(\rho_--\rho_+)^2}-1
    \,.
\end{equation}
Regarding the rhs, we must calculate the relative entropy between the unitarly evolved NESS and the NESS:
\begin{equation}
\begin{split}
    \Sigma_{rel} &= \text{S} (U\rho_{\text{ness}} U^\dagger||\rho_{\text{ness}})\\ &=(\rho_+-\rho_-)\text{ln}\left(\frac{\rho_+}{\rho_-}\right)\,,
\end{split}
\end{equation}
definitely getting for the rhs
\begin{equation}
    \frac{2}{e^{\Sigma_{rel}}-1}=\frac{2}{\Big({\frac{\rho_+}{\rho_-}}\Big)^{(\rho_+-\rho_-)}-1}\,.
\end{equation}
Hence, we need to prove that (see also \cite{Proesmans_2017} for an analogous proof but in a classical context)
\begin{equation}
    \frac{\rho_-+\rho_+}{(\rho_--\rho_+)^2}-1  \geq  \frac{2}{\Big({\frac{\rho_-}{\rho_+}}\Big)^{\rho_--\rho_+}-1}\,.
\end{equation}
In order to do so,
we make the change of variables $x = \rho_--\rho_+$ and $y={\rho_-}/{\rho_+}$. Therefore, $1\geq x \geq 0$ and $y\geq1$.
We can now rewrite the inequality as
\begin{equation}
    \frac{y+1}{x(y-1)}-1 \geq \frac{2}{y^x-1} 
    \,.
\end{equation}
We make a new change of variables $z=y^x \leq y$ so $x=\ln{z}/\ln{y}$ and we can rewrite the inequality as
\begin{equation}
    \frac{(y+1)\ln{y}}{(y-1)\ln{z}}-1 \geq \frac{2}{z-1}\,,
\end{equation}
which is equivalent to
\begin{equation}
    \frac{(y+1)\ln{y}}{(y-1)\ln{z}}\geq\frac{z+1}{z-1}
\Leftrightarrow
    \frac{y+1}{y-1}\ln{y}\geq\frac{z+1}{z-1}\ln{z}
\end{equation}
This last line is true since $y\geq z \geq 1$ and the function $x \mapsto \frac{x+1}{x-1}\ln{x}$ is increasing for $x\geq 1$.
}

\section{Entanglement criteria based on thermodynamic quantities}
\label{app:ent-criterio}
Here we discuss further how in our framework it is immediate to build entanglement criteria based on thermodynamic quantities. 
The first example has been reported in Ineq.\,\eqref{thermo_ent-cond}.
From such inequality,
we can further deduce conditions on the adimensional work variance $v:=(\langle W^2\rangle-\langle W \rangle^2 )/(4 g^2)$ and the adimensional average work $w:=\langle W\rangle / (2g)=\rho_--\rho_+$:
\begin{equation}
\frac{2-3 w^2-\sqrt{6 \gamma -2} }{3} < v < \frac{2-3 w^2+\sqrt{6 \gamma -2} }{3}~.
\label{ad-work-var}
\end{equation}
This means that, when entanglement is present, the higher the purity the wider the allowed range of $v$ and the higher the adimensional average work $w$ the lower the offset of such range.
For the specific case treated in Fig.~\ref{fig:TURs}(a), however, in the separable phase what is broken is the lower bound (namely $v$ is lower than such bound) and $v$ remains pretty close to the lower bound (but greater) in the entangled phase.
In summary, the relations \eqref{thermo_ent-cond} and \eqref{ad-work-var}, despite not playing a significant role in understanding the improvement due to entanglement, 
are totally equivalent to \eqref{ent-cond}. 
This implies that they represent entanglement criteria which are accessible through the purity of the NESS and, more importantly, through thermodynamic quantities.\\

\section{Finding the closest separable state}
\label{Appendix-minimization}
Following Ref.\,\cite{Verstraete2002}, here we show how we can find the closest separable state in the Frobenius norm, namely the argument of the minimization \eqref{minimization-main}, in formulas
\begin{eqnarray}
\bar{\rho}_{\text{sep}}= {\rm arg}\left(
\begin{array}{rrclcl}
\displaystyle \min_{\rho_{\text{sep}}} & \multicolumn{3}{l}{{\lVert \rho_{\text{ness}}-\rho_{\text{sep}}   \rVert}_2}\\
\textrm{s.t.} 
& \rho_{\text{sep}} \geq 0,\,{\rm Tr}(\rho_{\text{sep}})=1\\
& \rho_{\text{sep}}^{PT}\geq 0 
\label{minimization}
\end{array}
\quad\qquad
\right)\,.
\end{eqnarray}
The Frobenius norm, defined as 
\begin{equation}
{\lVert A   \rVert}_2^2:=\sum_{ij} \lvert A_{ij} \rvert^2\,,
\end{equation}
is invariant under partial transposition.
Hence the minimization in \eqref{minimization} is equivalent to
\begin{equation}
\begin{array}{rrclcl}
\displaystyle \min_{\rho_{\text{sep}}} & \multicolumn{3}{l}{{\lVert \rho_{\text{ness}}^{PT}-\rho_{\text{sep}}^{PT}   \rVert}_2}\\
\textrm{s.t.} 
& \rho_{\text{sep}} \geq 0,\,{\rm Tr}(\rho_{\text{sep}})=1\\
& \rho_{\text{sep}}^{PT}\geq 0\,. 
\label{minimization-2}
\end{array}
\end{equation}
where $\rho_{\text{ness}}^{PT}$ is a non-positive operator (iff the state $\rho_{\text{ness}}$ is entangled in the case of two qubits)
and $\rho_{\text{sep}}^{PT}$ is a proper density operator whose partially transpose state is positive.
Therefore \eqref{minimization-2} is equivalent to
\begin{equation}
\begin{array}{rrclcl}
\displaystyle \min_{\sigma} & \multicolumn{3}{l}{{\lVert \rho_{\text{ness}}^{PT}-\sigma  \rVert}_2}\\
\textrm{s.t.} 
& \sigma \geq 0,\,{\rm Tr}(\sigma)=1\\
& \sigma^{PT}\geq 0\,. 
\label{minimization-3}
\end{array}
\end{equation}
The minimization \eqref{minimization-3} returns an objective function that is always greater or equal than 
\begin{equation}
\begin{array}{rrclcl}
\displaystyle \min_{\sigma} & \multicolumn{3}{l}{{\lVert \rho_{\text{ness}}^{PT}-\sigma  \rVert}_2}\\
\textrm{s.t.} 
& \sigma \geq 0,\,{\rm Tr}(\sigma)=1
\label{minimization-4}
\end{array}
\end{equation}
namely the one without the constraint 
$\sigma^{PT}\geq 0$.
\eqref{minimization-4} is hence a relaxation of \eqref{minimization-3}.
To solve \eqref{minimization-4} one can apply the analytical method of Refs.\,\cite{Verstraete2002,smolin2012mle} to find the minimum $\bar{\sigma}= arg $ \eqref{minimization-4}, and then check by inspection if $\bar{\sigma}^{PT}\geq 0$. 
If $\bar{\sigma}^{PT}\geq 0$ one can identify
$\bar{\rho}_{\text{sep}}=\bar{\sigma}^{PT}$.
The relation
$\bar{\sigma}^{PT}\geq 0$ was proven by inspection to be satisfied for the examples we have provided
in the insets of Fig.~\ref{fig:TURs} and Fig.~\ref{fig:TUR-breaking}.
\\
Finally, we report in Fig.~\ref{fig:workvariance},
for the unitary \eqref{swapent}, the changes in the squared average work  and in the work variance obtained by replacing the NESS with its closest separable state.
We notice that such replacement implies that 
the average work decreases while the variance remains almost unchanged, leading to an increase in the relative error.

\section{\label{app:randU} Unitary of violation} 
We report the unitary matrix that allows one to obtain the violation plotted in Fig.~\ref{fig:TUR-breaking}:
\begin{widetext}
{
\begin{equation}
\label{Uviol}
U_{viol}=
    \begin{pmatrix}
    0.61214-0.084476 i &  0.442141-0.20187 i &  0.197476-0.549142 i &  0.166498-0.116762 i\\ -0.000772-0.210944 i &  0.125315+0.662622 i &  -0.440385-0.240318 i &  0.347386+0.358276 i\\ 0.250147-0.159182 i &  0.198848+0.471307 i &  -0.095917+0.135918 i &  -0.678087-0.40366 i\\ -0.691565+0.086468 i &  0.190366+0.105252 i &  0.175807-0.590908 i &  -0.133258-0.262879 i
    \end{pmatrix}
    \,,
\end{equation}
 }
\end{widetext}
having chosen the representation in the basis $(\ket{11}, \ket{10}, \ket{01}, \ket{00})$ and having rounded the entries to 6 decimals.

\end{appendix}

\bibliography{apssamp}

\providecommand{\noopsort}[1]{}\providecommand{\singleletter}[1]{#1}%
\begin{thebibliography}{45}%
\makeatletter
\providecommand \@ifxundefined [1]{%
 \@ifx{#1\undefined}
}%
\providecommand \@ifnum [1]{%
 \ifnum #1\expandafter \@firstoftwo
 \else \expandafter \@secondoftwo
 \fi
}%
\providecommand \@ifx [1]{%
 \ifx #1\expandafter \@firstoftwo
 \else \expandafter \@secondoftwo
 \fi
}%
\providecommand \natexlab [1]{#1}%
\providecommand \enquote  [1]{``#1''}%
\providecommand \bibnamefont  [1]{#1}%
\providecommand \bibfnamefont [1]{#1}%
\providecommand \citenamefont [1]{#1}%
\providecommand \href@noop [0]{\@secondoftwo}%
\providecommand \href [0]{\begingroup \@sanitize@url \@href}%
\providecommand \@href[1]{\@@startlink{#1}\@@href}%
\providecommand \@@href[1]{\endgroup#1\@@endlink}%
\providecommand \@sanitize@url [0]{\catcode `\\12\catcode `\$12\catcode
  `\&12\catcode `\#12\catcode `\^12\catcode `\_12\catcode `\%12\relax}%
\providecommand \@@startlink[1]{}%
\providecommand \@@endlink[0]{}%
\providecommand \url  [0]{\begingroup\@sanitize@url \@url }%
\providecommand \@url [1]{\endgroup\@href {#1}{\urlprefix }}%
\providecommand \urlprefix  [0]{URL }%
\providecommand \Eprint [0]{\href }%
\providecommand \doibase [0]{http://dx.doi.org/}%
\providecommand \selectlanguage [0]{\@gobble}%
\providecommand \bibinfo  [0]{\@secondoftwo}%
\providecommand \bibfield  [0]{\@secondoftwo}%
\providecommand \translation [1]{[#1]}%
\providecommand \BibitemOpen [0]{}%
\providecommand \bibitemStop [0]{}%
\providecommand \bibitemNoStop [0]{.\EOS\space}%
\providecommand \EOS [0]{\spacefactor3000\relax}%
\providecommand \BibitemShut  [1]{\csname bibitem#1\endcsname}%
\let\auto@bib@innerbib\@empty
\bibitem [{\citenamefont {Vinjanampathy}\ and\ \citenamefont
  {Anders}(2016)}]{anders2016reviewqtd}%
  \BibitemOpen
  \bibfield  {author} {\bibinfo {author} {\bibfnamefont {S.}~\bibnamefont
  {Vinjanampathy}}\ and\ \bibinfo {author} {\bibfnamefont {J.}~\bibnamefont
  {Anders}},\ }\href@noop {} {\bibfield  {journal} {\bibinfo  {journal}
  {Contemporary Physics}\ }\textbf {\bibinfo {volume} {57}},\ \bibinfo {pages}
  {545} (\bibinfo {year} {2016})}\BibitemShut {NoStop}%
\bibitem [{\citenamefont {Campisi}\ \emph {et~al.}(2011)\citenamefont
  {Campisi}, \citenamefont {H\"anggi},\ and\ \citenamefont
  {Talkner}}]{campisi2011rmp}%
  \BibitemOpen
  \bibfield  {author} {\bibinfo {author} {\bibfnamefont {M.}~\bibnamefont
  {Campisi}}, \bibinfo {author} {\bibfnamefont {P.}~\bibnamefont {H\"anggi}}, \
  and\ \bibinfo {author} {\bibfnamefont {P.}~\bibnamefont {Talkner}},\ }\href
  {\doibase 10.1103/RevModPhys.83.771} {\bibfield  {journal} {\bibinfo
  {journal} {Rev. Mod. Phys.}\ }\textbf {\bibinfo {volume} {83}},\ \bibinfo
  {pages} {771} (\bibinfo {year} {2011})}\BibitemShut {NoStop}%
\bibitem [{\citenamefont {Alicki}\ and\ \citenamefont
  {Kosloff}(2018)}]{alicki2018introduction}%
  \BibitemOpen
  \bibfield  {author} {\bibinfo {author} {\bibfnamefont {R.}~\bibnamefont
  {Alicki}}\ and\ \bibinfo {author} {\bibfnamefont {R.}~\bibnamefont
  {Kosloff}},\ }in\ \href@noop {} {\emph {\bibinfo {booktitle} {Thermodynamics
  in the Quantum Regime}}}\ (\bibinfo  {publisher} {Springer},\ \bibinfo {year}
  {2018})\ pp.\ \bibinfo {pages} {1--33}\BibitemShut {NoStop}%
\bibitem [{\citenamefont {Auff\`eves}(2022)}]{auffeves2022initiative}%
  \BibitemOpen
  \bibfield  {author} {\bibinfo {author} {\bibfnamefont {A.}~\bibnamefont
  {Auff\`eves}},\ }\href {\doibase 10.1103/PRXQuantum.3.020101} {\bibfield
  {journal} {\bibinfo  {journal} {PRX Quantum}\ }\textbf {\bibinfo {volume}
  {3}},\ \bibinfo {pages} {020101} (\bibinfo {year} {2022})}\BibitemShut
  {NoStop}%
\bibitem [{\citenamefont {Barato}\ and\ \citenamefont
  {Seifert}(2015)}]{baratoseifert2015tur}%
  \BibitemOpen
  \bibfield  {author} {\bibinfo {author} {\bibfnamefont {A.~C.}\ \bibnamefont
  {Barato}}\ and\ \bibinfo {author} {\bibfnamefont {U.}~\bibnamefont
  {Seifert}},\ }\href {\doibase 10.1103/PhysRevLett.114.158101} {\bibfield
  {journal} {\bibinfo  {journal} {Phys. Rev. Lett.}\ }\textbf {\bibinfo
  {volume} {114}},\ \bibinfo {pages} {158101} (\bibinfo {year}
  {2015})}\BibitemShut {NoStop}%
\bibitem [{\citenamefont {Pietzonka}\ \emph {et~al.}(2016)\citenamefont
  {Pietzonka}, \citenamefont {Barato},\ and\ \citenamefont
  {Seifert}}]{seifert2016pre}%
  \BibitemOpen
  \bibfield  {author} {\bibinfo {author} {\bibfnamefont {P.}~\bibnamefont
  {Pietzonka}}, \bibinfo {author} {\bibfnamefont {A.~C.}\ \bibnamefont
  {Barato}}, \ and\ \bibinfo {author} {\bibfnamefont {U.}~\bibnamefont
  {Seifert}},\ }\href {\doibase 10.1103/PhysRevE.93.052145} {\bibfield
  {journal} {\bibinfo  {journal} {Phys. Rev. E}\ }\textbf {\bibinfo {volume}
  {93}},\ \bibinfo {pages} {052145} (\bibinfo {year} {2016})}\BibitemShut
  {NoStop}%
\bibitem [{\citenamefont {Pietzonka}\ \emph {et~al.}(2017)\citenamefont
  {Pietzonka}, \citenamefont {Ritort},\ and\ \citenamefont
  {Seifert}}]{seifert2017pre}%
  \BibitemOpen
  \bibfield  {author} {\bibinfo {author} {\bibfnamefont {P.}~\bibnamefont
  {Pietzonka}}, \bibinfo {author} {\bibfnamefont {F.}~\bibnamefont {Ritort}}, \
  and\ \bibinfo {author} {\bibfnamefont {U.}~\bibnamefont {Seifert}},\ }\href
  {\doibase 10.1103/PhysRevE.96.012101} {\bibfield  {journal} {\bibinfo
  {journal} {Phys. Rev. E}\ }\textbf {\bibinfo {volume} {96}},\ \bibinfo
  {pages} {012101} (\bibinfo {year} {2017})}\BibitemShut {NoStop}%
\bibitem [{\citenamefont {Pietzonka}\ and\ \citenamefont
  {Seifert}(2018)}]{seifert2018steady}%
  \BibitemOpen
  \bibfield  {author} {\bibinfo {author} {\bibfnamefont {P.}~\bibnamefont
  {Pietzonka}}\ and\ \bibinfo {author} {\bibfnamefont {U.}~\bibnamefont
  {Seifert}},\ }\href {\doibase 10.1103/PhysRevLett.120.190602} {\bibfield
  {journal} {\bibinfo  {journal} {Phys. Rev. Lett.}\ }\textbf {\bibinfo
  {volume} {120}},\ \bibinfo {pages} {190602} (\bibinfo {year}
  {2018})}\BibitemShut {NoStop}%
\bibitem [{\citenamefont {Horowitz}\ and\ \citenamefont
  {Gingrich}(2020)}]{horowitz2020thermUR}%
  \BibitemOpen
  \bibfield  {author} {\bibinfo {author} {\bibfnamefont {J.~M.}\ \bibnamefont
  {Horowitz}}\ and\ \bibinfo {author} {\bibfnamefont {T.~R.}\ \bibnamefont
  {Gingrich}},\ }\href {\doibase 10.1038/s41567-019-0702-6} {\bibfield
  {journal} {\bibinfo  {journal} {Nature Physics}\ }\textbf {\bibinfo {volume}
  {16}},\ \bibinfo {pages} {15} (\bibinfo {year} {2020})}\BibitemShut {NoStop}%
\bibitem [{\citenamefont {Timpanaro}\ \emph {et~al.}(2019)\citenamefont
  {Timpanaro}, \citenamefont {Guarnieri}, \citenamefont {Goold},\ and\
  \citenamefont {Landi}}]{landi2019tur}%
  \BibitemOpen
  \bibfield  {author} {\bibinfo {author} {\bibfnamefont {A.~M.}\ \bibnamefont
  {Timpanaro}}, \bibinfo {author} {\bibfnamefont {G.}~\bibnamefont
  {Guarnieri}}, \bibinfo {author} {\bibfnamefont {J.}~\bibnamefont {Goold}}, \
  and\ \bibinfo {author} {\bibfnamefont {G.~T.}\ \bibnamefont {Landi}},\ }\href
  {\doibase 10.1103/PhysRevLett.123.090604} {\bibfield  {journal} {\bibinfo
  {journal} {Phys. Rev. Lett.}\ }\textbf {\bibinfo {volume} {123}},\ \bibinfo
  {pages} {090604} (\bibinfo {year} {2019})}\BibitemShut {NoStop}%
\bibitem [{\citenamefont {Jarzynski}\ and\ \citenamefont
  {W\'ojcik}(2004)}]{jarzynski2004exchange}%
  \BibitemOpen
  \bibfield  {author} {\bibinfo {author} {\bibfnamefont {C.}~\bibnamefont
  {Jarzynski}}\ and\ \bibinfo {author} {\bibfnamefont {D.~K.}\ \bibnamefont
  {W\'ojcik}},\ }\href {\doibase 10.1103/PhysRevLett.92.230602} {\bibfield
  {journal} {\bibinfo  {journal} {Phys. Rev. Lett.}\ }\textbf {\bibinfo
  {volume} {92}},\ \bibinfo {pages} {230602} (\bibinfo {year}
  {2004})}\BibitemShut {NoStop}%
\bibitem [{avg()}]{avgEP}%
  \BibitemOpen
  \href@noop {} {\bibinfo  {journal} {From now on, for brevity, we will often
  call it {\it entropy production}, dropping the term {\it average}}\
  }\BibitemShut {NoStop}%
\bibitem [{\citenamefont {Cappellaro}\ \emph {et~al.}(2005)\citenamefont
  {Cappellaro}, \citenamefont {Emerson}, \citenamefont {Boulant}, \citenamefont
  {Ramanathan}, \citenamefont {Lloyd},\ and\ \citenamefont
  {Cory}}]{cappellaro2005entanglement}%
  \BibitemOpen
\bibfield  {journal} {  }\bibfield  {author} {\bibinfo {author} {\bibfnamefont
  {P.}~\bibnamefont {Cappellaro}}, \bibinfo {author} {\bibfnamefont
  {J.}~\bibnamefont {Emerson}}, \bibinfo {author} {\bibfnamefont
  {N.}~\bibnamefont {Boulant}}, \bibinfo {author} {\bibfnamefont
  {C.}~\bibnamefont {Ramanathan}}, \bibinfo {author} {\bibfnamefont
  {S.}~\bibnamefont {Lloyd}}, \ and\ \bibinfo {author} {\bibfnamefont {D.~G.}\
  \bibnamefont {Cory}},\ }\href {\doibase 10.1103/PhysRevLett.94.020502}
  {\bibfield  {journal} {\bibinfo  {journal} {Phys. Rev. Lett.}\ }\textbf
  {\bibinfo {volume} {94}},\ \bibinfo {pages} {020502} (\bibinfo {year}
  {2005})}\BibitemShut {NoStop}%
\bibitem [{\citenamefont {T\'oth}(2012)}]{toth2012multipartite}%
  \BibitemOpen
  \bibfield  {author} {\bibinfo {author} {\bibfnamefont {G.}~\bibnamefont
  {T\'oth}},\ }\href {\doibase 10.1103/PhysRevA.85.022322} {\bibfield
  {journal} {\bibinfo  {journal} {Phys. Rev. A}\ }\textbf {\bibinfo {volume}
  {85}},\ \bibinfo {pages} {022322} (\bibinfo {year} {2012})}\BibitemShut
  {NoStop}%
\bibitem [{\citenamefont {Horowitz}\ and\ \citenamefont
  {Gingrich}(2017)}]{Horowitz2017rapidcomm}%
  \BibitemOpen
  \bibfield  {author} {\bibinfo {author} {\bibfnamefont {J.~M.}\ \bibnamefont
  {Horowitz}}\ and\ \bibinfo {author} {\bibfnamefont {T.~R.}\ \bibnamefont
  {Gingrich}},\ }\href {\doibase 10.1103/PhysRevE.96.020103} {\bibfield
  {journal} {\bibinfo  {journal} {Phys. Rev. E}\ }\textbf {\bibinfo {volume}
  {96}},\ \bibinfo {pages} {020103} (\bibinfo {year} {2017})}\BibitemShut
  {NoStop}%
\bibitem [{\citenamefont {Ptaszy\ifmmode~\acute{n}\else
  \'{n}\fi{}ski}(2018)}]{2018-Coherence-qdots-ness-tur}%
  \BibitemOpen
  \bibfield  {author} {\bibinfo {author} {\bibfnamefont {K.}~\bibnamefont
  {Ptaszy\ifmmode~\acute{n}\else \'{n}\fi{}ski}},\ }\href {\doibase
  10.1103/PhysRevB.98.085425} {\bibfield  {journal} {\bibinfo  {journal} {Phys.
  Rev. B}\ }\textbf {\bibinfo {volume} {98}},\ \bibinfo {pages} {085425}
  (\bibinfo {year} {2018})}\BibitemShut {NoStop}%
\bibitem [{\citenamefont {Guarnieri}\ \emph {et~al.}(2019)\citenamefont
  {Guarnieri}, \citenamefont {Landi}, \citenamefont {Clark},\ and\
  \citenamefont {Goold}}]{guarnieri2019ness}%
  \BibitemOpen
  \bibfield  {author} {\bibinfo {author} {\bibfnamefont {G.}~\bibnamefont
  {Guarnieri}}, \bibinfo {author} {\bibfnamefont {G.~T.}\ \bibnamefont
  {Landi}}, \bibinfo {author} {\bibfnamefont {S.~R.}\ \bibnamefont {Clark}}, \
  and\ \bibinfo {author} {\bibfnamefont {J.}~\bibnamefont {Goold}},\ }\href
  {\doibase 10.1103/PhysRevResearch.1.033021} {\bibfield  {journal} {\bibinfo
  {journal} {Phys. Rev. Research}\ }\textbf {\bibinfo {volume} {1}},\ \bibinfo
  {pages} {033021} (\bibinfo {year} {2019})}\BibitemShut {NoStop}%
\bibitem [{\citenamefont {Sacchi}(2021)}]{sacchi2021qudits}%
  \BibitemOpen
  \bibfield  {author} {\bibinfo {author} {\bibfnamefont {M.~F.}\ \bibnamefont
  {Sacchi}},\ }\href {\doibase 10.1103/PhysRevA.104.012217} {\bibfield
  {journal} {\bibinfo  {journal} {Phys. Rev. A}\ }\textbf {\bibinfo {volume}
  {104}},\ \bibinfo {pages} {012217} (\bibinfo {year} {2021})}\BibitemShut
  {NoStop}%
\bibitem [{\citenamefont {Hofer}\ \emph {et~al.}(2017)\citenamefont {Hofer},
  \citenamefont {Perarnau-Llobet}, \citenamefont {Miranda}, \citenamefont
  {Haack}, \citenamefont {Silva}, \citenamefont {Brask},\ and\ \citenamefont
  {Brunner}}]{Hofer_2017}%
  \BibitemOpen
  \bibfield  {author} {\bibinfo {author} {\bibfnamefont {P.~P.}\ \bibnamefont
  {Hofer}}, \bibinfo {author} {\bibfnamefont {M.}~\bibnamefont
  {Perarnau-Llobet}}, \bibinfo {author} {\bibfnamefont {L.~D.~M.}\ \bibnamefont
  {Miranda}}, \bibinfo {author} {\bibfnamefont {G.}~\bibnamefont {Haack}},
  \bibinfo {author} {\bibfnamefont {R.}~\bibnamefont {Silva}}, \bibinfo
  {author} {\bibfnamefont {J.~B.}\ \bibnamefont {Brask}}, \ and\ \bibinfo
  {author} {\bibfnamefont {N.}~\bibnamefont {Brunner}},\ }\href {\doibase
  10.1088/1367-2630/aa964f} {\bibfield  {journal} {\bibinfo  {journal} {New
  Journal of Physics}\ }\textbf {\bibinfo {volume} {19}},\ \bibinfo {pages}
  {123037} (\bibinfo {year} {2017})}\BibitemShut {NoStop}%
\bibitem [{\citenamefont {Khandelwal}\ \emph {et~al.}(2020)\citenamefont
  {Khandelwal}, \citenamefont {Palazzo}, \citenamefont {Brunner},\ and\
  \citenamefont {Haack}}]{Khandelwal_2020}%
  \BibitemOpen
  \bibfield  {author} {\bibinfo {author} {\bibfnamefont {S.}~\bibnamefont
  {Khandelwal}}, \bibinfo {author} {\bibfnamefont {N.}~\bibnamefont {Palazzo}},
  \bibinfo {author} {\bibfnamefont {N.}~\bibnamefont {Brunner}}, \ and\
  \bibinfo {author} {\bibfnamefont {G.}~\bibnamefont {Haack}},\ }\href
  {\doibase 10.1088/1367-2630/ab9983} {\bibfield  {journal} {\bibinfo
  {journal} {New Journal of Physics}\ }\textbf {\bibinfo {volume} {22}},\
  \bibinfo {pages} {073039} (\bibinfo {year} {2020})}\BibitemShut {NoStop}%
\bibitem [{\citenamefont {Gorini}\ \emph {et~al.}(1976)\citenamefont {Gorini},
  \citenamefont {Kossakowski},\ and\ \citenamefont
  {Sudarshan}}]{gorini1976completely}%
  \BibitemOpen
  \bibfield  {author} {\bibinfo {author} {\bibfnamefont {V.}~\bibnamefont
  {Gorini}}, \bibinfo {author} {\bibfnamefont {A.}~\bibnamefont {Kossakowski}},
  \ and\ \bibinfo {author} {\bibfnamefont {E.~C.~G.}\ \bibnamefont
  {Sudarshan}},\ }\href {\doibase 10.1063/1.522979} {\bibfield  {journal}
  {\bibinfo  {journal} {Journal of Mathematical Physics}\ }\textbf {\bibinfo
  {volume} {17}},\ \bibinfo {pages} {821} (\bibinfo {year} {1976})},\ \Eprint
  {http://arxiv.org/abs/https://aip.scitation.org/doi/pdf/10.1063/1.522979}
  {https://aip.scitation.org/doi/pdf/10.1063/1.522979} \BibitemShut {NoStop}%
\bibitem [{\citenamefont {Lindblad}(1976)}]{lindblad1976generators}%
  \BibitemOpen
  \bibfield  {author} {\bibinfo {author} {\bibfnamefont {G.}~\bibnamefont
  {Lindblad}},\ }\href {\doibase 10.1007/BF01608499} {\bibfield  {journal}
  {\bibinfo  {journal} {Communications in Mathematical Physics}\ }\textbf
  {\bibinfo {volume} {48}},\ \bibinfo {pages} {119} (\bibinfo {year}
  {1976})}\BibitemShut {NoStop}%
\bibitem [{\citenamefont {Gonz\'{a}lez}\ \emph {et~al.}(2017)\citenamefont
  {Gonz\'{a}lez}, \citenamefont {Correa}, \citenamefont {Nocerino},
  \citenamefont {Palao}, \citenamefont {Alonso},\ and\ \citenamefont
  {Adesso}}]{adesso2017locvsglob}%
  \BibitemOpen
  \bibfield  {author} {\bibinfo {author} {\bibfnamefont {J.~O.}\ \bibnamefont
  {Gonz\'{a}lez}}, \bibinfo {author} {\bibfnamefont {L.~A.}\ \bibnamefont
  {Correa}}, \bibinfo {author} {\bibfnamefont {G.}~\bibnamefont {Nocerino}},
  \bibinfo {author} {\bibfnamefont {J.~P.}\ \bibnamefont {Palao}}, \bibinfo
  {author} {\bibfnamefont {D.}~\bibnamefont {Alonso}}, \ and\ \bibinfo {author}
  {\bibfnamefont {G.}~\bibnamefont {Adesso}},\ }\href {\doibase
  10.1142/S1230161217400108} {\bibfield  {journal} {\bibinfo  {journal} {Open
  Systems \& Information Dynamics}\ }\textbf {\bibinfo {volume} {24}},\
  \bibinfo {pages} {1740010} (\bibinfo {year} {2017})},\ \Eprint
  {http://arxiv.org/abs/https://doi.org/10.1142/S1230161217400108}
  {https://doi.org/10.1142/S1230161217400108} \BibitemShut {NoStop}%
\bibitem [{\citenamefont {Alicki}(1976)}]{ALICKI1976249}%
  \BibitemOpen
  \bibfield  {author} {\bibinfo {author} {\bibfnamefont {R.}~\bibnamefont
  {Alicki}},\ }\href {\doibase https://doi.org/10.1016/0034-4877(76)90046-X}
  {\bibfield  {journal} {\bibinfo  {journal} {Reports on Mathematical Physics}\
  }\textbf {\bibinfo {volume} {10}},\ \bibinfo {pages} {249} (\bibinfo {year}
  {1976})}\BibitemShut {NoStop}%
\bibitem [{\citenamefont {Spohn}(1978{\natexlab{a}})}]{spohn1978semigroup}%
  \BibitemOpen
  \bibfield  {author} {\bibinfo {author} {\bibfnamefont {H.}~\bibnamefont
  {Spohn}},\ }\href {\doibase 10.1063/1.523789} {\bibfield  {journal} {\bibinfo
   {journal} {Journal of Mathematical Physics}\ }\textbf {\bibinfo {volume}
  {19}},\ \bibinfo {pages} {1227} (\bibinfo {year} {1978}{\natexlab{a}})},\
  \Eprint {http://arxiv.org/abs/https://doi.org/10.1063/1.523789}
  {https://doi.org/10.1063/1.523789} \BibitemShut {NoStop}%
\bibitem [{\citenamefont {Tasaki}(2000)}]{tasaki2000jarzynski}%
  \BibitemOpen
  \bibfield  {author} {\bibinfo {author} {\bibfnamefont {H.}~\bibnamefont
  {Tasaki}},\ }\href@noop {} {\bibfield  {journal} {\bibinfo  {journal} {arXiv
  preprint cond-mat/0009244}\ } (\bibinfo {year} {2000})}\BibitemShut {NoStop}%
\bibitem [{\citenamefont {Kurchan}(2000)}]{kurchan2000quantum}%
  \BibitemOpen
  \bibfield  {author} {\bibinfo {author} {\bibfnamefont {J.}~\bibnamefont
  {Kurchan}},\ }\href@noop {} {\bibfield  {journal} {\bibinfo  {journal} {arXiv
  preprint cond-mat/0007360}\ } (\bibinfo {year} {2000})}\BibitemShut {NoStop}%
\bibitem [{\citenamefont {Mukamel}(2003)}]{Mukamel2000prlqjarz}%
  \BibitemOpen
  \bibfield  {author} {\bibinfo {author} {\bibfnamefont {S.}~\bibnamefont
  {Mukamel}},\ }\href {\doibase 10.1103/PhysRevLett.90.170604} {\bibfield
  {journal} {\bibinfo  {journal} {Phys. Rev. Lett.}\ }\textbf {\bibinfo
  {volume} {90}},\ \bibinfo {pages} {170604} (\bibinfo {year}
  {2003})}\BibitemShut {NoStop}%
\bibitem [{\citenamefont {Talkner}\ \emph {et~al.}(2007)\citenamefont
  {Talkner}, \citenamefont {Lutz},\ and\ \citenamefont
  {H\"anggi}}]{hanggi2007WnotObs}%
  \BibitemOpen
  \bibfield  {author} {\bibinfo {author} {\bibfnamefont {P.}~\bibnamefont
  {Talkner}}, \bibinfo {author} {\bibfnamefont {E.}~\bibnamefont {Lutz}}, \
  and\ \bibinfo {author} {\bibfnamefont {P.}~\bibnamefont {H\"anggi}},\ }\href
  {\doibase 10.1103/PhysRevE.75.050102} {\bibfield  {journal} {\bibinfo
  {journal} {Phys. Rev. E}\ }\textbf {\bibinfo {volume} {75}},\ \bibinfo
  {pages} {050102} (\bibinfo {year} {2007})}\BibitemShut {NoStop}%
\bibitem [{\citenamefont {Esposito}\ \emph {et~al.}(2010)\citenamefont
  {Esposito}, \citenamefont {Lindenberg},\ and\ \citenamefont {den
  Broeck}}]{Esposito_2010}%
  \BibitemOpen
  \bibfield  {author} {\bibinfo {author} {\bibfnamefont {M.}~\bibnamefont
  {Esposito}}, \bibinfo {author} {\bibfnamefont {K.}~\bibnamefont
  {Lindenberg}}, \ and\ \bibinfo {author} {\bibfnamefont {C.~V.}\ \bibnamefont
  {den Broeck}},\ }\href {\doibase 10.1088/1367-2630/12/1/013013} {\bibfield
  {journal} {\bibinfo  {journal} {New Journal of Physics}\ }\textbf {\bibinfo
  {volume} {12}},\ \bibinfo {pages} {013013} (\bibinfo {year}
  {2010})}\BibitemShut {NoStop}%
\bibitem [{\citenamefont {Barra}(2015)}]{barra2015thermodynamic}%
  \BibitemOpen
  \bibfield  {author} {\bibinfo {author} {\bibfnamefont {F.}~\bibnamefont
  {Barra}},\ }\href {\doibase 10.1038/srep14873} {\bibfield  {journal}
  {\bibinfo  {journal} {Scientific reports}\ }\textbf {\bibinfo {volume} {5}},\
  \bibinfo {pages} {1} (\bibinfo {year} {2015})}\BibitemShut {NoStop}%
\bibitem [{\citenamefont {Spohn}(1978{\natexlab{b}})}]{spohn1978entropy}%
  \BibitemOpen
  \bibfield  {author} {\bibinfo {author} {\bibfnamefont {H.}~\bibnamefont
  {Spohn}},\ }\href@noop {} {\bibfield  {journal} {\bibinfo  {journal} {Journal
  of Mathematical Physics}\ }\textbf {\bibinfo {volume} {19}},\ \bibinfo
  {pages} {1227} (\bibinfo {year} {1978}{\natexlab{b}})}\BibitemShut {NoStop}%
\bibitem [{\citenamefont {Manzano}\ \emph {et~al.}(2015)\citenamefont
  {Manzano}, \citenamefont {Horowitz},\ and\ \citenamefont
  {Parrondo}}]{PhysRevE.92.032129}%
  \BibitemOpen
  \bibfield  {author} {\bibinfo {author} {\bibfnamefont {G.}~\bibnamefont
  {Manzano}}, \bibinfo {author} {\bibfnamefont {J.~M.}\ \bibnamefont
  {Horowitz}}, \ and\ \bibinfo {author} {\bibfnamefont {J.~M.~R.}\ \bibnamefont
  {Parrondo}},\ }\href {\doibase 10.1103/PhysRevE.92.032129} {\bibfield
  {journal} {\bibinfo  {journal} {Phys. Rev. E}\ }\textbf {\bibinfo {volume}
  {92}},\ \bibinfo {pages} {032129} (\bibinfo {year} {2015})}\BibitemShut
  {NoStop}%
\bibitem [{\citenamefont {Manzano}\ \emph {et~al.}(2018)\citenamefont
  {Manzano}, \citenamefont {Horowitz},\ and\ \citenamefont
  {Parrondo}}]{PhysRevX.8.031037}%
  \BibitemOpen
  \bibfield  {author} {\bibinfo {author} {\bibfnamefont {G.}~\bibnamefont
  {Manzano}}, \bibinfo {author} {\bibfnamefont {J.~M.}\ \bibnamefont
  {Horowitz}}, \ and\ \bibinfo {author} {\bibfnamefont {J.~M.~R.}\ \bibnamefont
  {Parrondo}},\ }\href {\doibase 10.1103/PhysRevX.8.031037} {\bibfield
  {journal} {\bibinfo  {journal} {Phys. Rev. X}\ }\textbf {\bibinfo {volume}
  {8}},\ \bibinfo {pages} {031037} (\bibinfo {year} {2018})}\BibitemShut
  {NoStop}%
\bibitem [{joh()}]{johansson2012qutip}%
  \BibitemOpen
  \href@noop {} {\bibinfo  {journal} {We simulated the dynamics of the
  two-qubit system using the Qutip library, J. R. Johansson, P. D. Nation, and
  F. Nori: {\it QuTiP 2: A Python framework for the dynamics of open quantum
  systems}, Comp. Phys. Comm. 184, 1234 (2013)}\ }\BibitemShut {NoStop}%
\bibitem [{\citenamefont {Merhav}\ and\ \citenamefont
  {Kafri}(2010)}]{Merhav_2010}%
  \BibitemOpen
\bibfield  {journal} {  }\bibfield  {author} {\bibinfo {author} {\bibfnamefont
  {N.}~\bibnamefont {Merhav}}\ and\ \bibinfo {author} {\bibfnamefont
  {Y.}~\bibnamefont {Kafri}},\ }\href {\doibase
  10.1088/1742-5468/2010/12/p12022} {\bibfield  {journal} {\bibinfo  {journal}
  {Journal of Statistical Mechanics: Theory and Experiment}\ }\textbf {\bibinfo
  {volume} {2010}},\ \bibinfo {pages} {P12022} (\bibinfo {year}
  {2010})}\BibitemShut {NoStop}%
\bibitem [{\citenamefont {Proesmans}\ and\ \citenamefont {den
  Broeck}(2017)}]{Proesmans_2017}%
  \BibitemOpen
  \bibfield  {author} {\bibinfo {author} {\bibfnamefont {K.}~\bibnamefont
  {Proesmans}}\ and\ \bibinfo {author} {\bibfnamefont {C.~V.}\ \bibnamefont
  {den Broeck}},\ }\href {\doibase 10.1209/0295-5075/119/20001} {\bibfield
  {journal} {\bibinfo  {journal} {{EPL} (Europhysics Letters)}\ }\textbf
  {\bibinfo {volume} {119}},\ \bibinfo {pages} {20001} (\bibinfo {year}
  {2017})}\BibitemShut {NoStop}%
\bibitem [{\citenamefont {Hasegawa}\ and\ \citenamefont
  {Van~Vu}(2019)}]{hasegawa2019fe}%
  \BibitemOpen
  \bibfield  {author} {\bibinfo {author} {\bibfnamefont {Y.}~\bibnamefont
  {Hasegawa}}\ and\ \bibinfo {author} {\bibfnamefont {T.}~\bibnamefont
  {Van~Vu}},\ }\href {\doibase 10.1103/PhysRevLett.123.110602} {\bibfield
  {journal} {\bibinfo  {journal} {Phys. Rev. Lett.}\ }\textbf {\bibinfo
  {volume} {123}},\ \bibinfo {pages} {110602} (\bibinfo {year}
  {2019})}\BibitemShut {NoStop}%
\bibitem [{\citenamefont {Agarwalla}\ and\ \citenamefont
  {Segal}(2018)}]{agarwalla2018assessing}%
  \BibitemOpen
  \bibfield  {author} {\bibinfo {author} {\bibfnamefont {B.~K.}\ \bibnamefont
  {Agarwalla}}\ and\ \bibinfo {author} {\bibfnamefont {D.}~\bibnamefont
  {Segal}},\ }\href {\doibase 10.1103/PhysRevB.98.155438} {\bibfield  {journal}
  {\bibinfo  {journal} {Phys. Rev. B}\ }\textbf {\bibinfo {volume} {98}},\
  \bibinfo {pages} {155438} (\bibinfo {year} {2018})}\BibitemShut {NoStop}%
\bibitem [{\citenamefont {Cangemi}\ \emph {et~al.}(2020)\citenamefont
  {Cangemi}, \citenamefont {Cataudella}, \citenamefont {Benenti}, \citenamefont
  {Sassetti},\ and\ \citenamefont {De~Filippis}}]{cangemi2020violation}%
  \BibitemOpen
  \bibfield  {author} {\bibinfo {author} {\bibfnamefont {L.~M.}\ \bibnamefont
  {Cangemi}}, \bibinfo {author} {\bibfnamefont {V.}~\bibnamefont {Cataudella}},
  \bibinfo {author} {\bibfnamefont {G.}~\bibnamefont {Benenti}}, \bibinfo
  {author} {\bibfnamefont {M.}~\bibnamefont {Sassetti}}, \ and\ \bibinfo
  {author} {\bibfnamefont {G.}~\bibnamefont {De~Filippis}},\ }\href {\doibase
  10.1103/PhysRevB.102.165418} {\bibfield  {journal} {\bibinfo  {journal}
  {Phys. Rev. B}\ }\textbf {\bibinfo {volume} {102}},\ \bibinfo {pages}
  {165418} (\bibinfo {year} {2020})}\BibitemShut {NoStop}%
\bibitem [{\citenamefont {Paneru}\ \emph {et~al.}(2020)\citenamefont {Paneru},
  \citenamefont {Dutta}, \citenamefont {Tlusty},\ and\ \citenamefont
  {Pak}}]{paneru2020reaching}%
  \BibitemOpen
  \bibfield  {author} {\bibinfo {author} {\bibfnamefont {G.}~\bibnamefont
  {Paneru}}, \bibinfo {author} {\bibfnamefont {S.}~\bibnamefont {Dutta}},
  \bibinfo {author} {\bibfnamefont {T.}~\bibnamefont {Tlusty}}, \ and\ \bibinfo
  {author} {\bibfnamefont {H.~K.}\ \bibnamefont {Pak}},\ }\href {\doibase
  10.1103/PhysRevE.102.032126} {\bibfield  {journal} {\bibinfo  {journal}
  {Phys. Rev. E}\ }\textbf {\bibinfo {volume} {102}},\ \bibinfo {pages}
  {032126} (\bibinfo {year} {2020})}\BibitemShut {NoStop}%
\bibitem [{\citenamefont {Verstraete}\ \emph {et~al.}(2002)\citenamefont
  {Verstraete}, \citenamefont {Dehaene},\ and\ \citenamefont
  {Moor}}]{Verstraete2002}%
  \BibitemOpen
  \bibfield  {author} {\bibinfo {author} {\bibfnamefont {F.}~\bibnamefont
  {Verstraete}}, \bibinfo {author} {\bibfnamefont {J.}~\bibnamefont {Dehaene}},
  \ and\ \bibinfo {author} {\bibfnamefont {B.~D.}\ \bibnamefont {Moor}},\
  }\href {\doibase 10.1080/09500340110115488} {\bibfield  {journal} {\bibinfo
  {journal} {Journal of Modern Optics}\ }\textbf {\bibinfo {volume} {49}},\
  \bibinfo {pages} {1277} (\bibinfo {year} {2002})},\ \Eprint
  {http://arxiv.org/abs/https://doi.org/10.1080/09500340110115488}
  {https://doi.org/10.1080/09500340110115488} \BibitemShut {NoStop}%
\bibitem [{\citenamefont {Van~Vu}\ and\ \citenamefont
  {Saito}(2022)}]{PhysRevLett128140602}%
  \BibitemOpen
  \bibfield  {author} {\bibinfo {author} {\bibfnamefont {T.}~\bibnamefont
  {Van~Vu}}\ and\ \bibinfo {author} {\bibfnamefont {K.}~\bibnamefont {Saito}},\
  }\href {\doibase 10.1103/PhysRevLett.128.140602} {\bibfield  {journal}
  {\bibinfo  {journal} {Phys. Rev. Lett.}\ }\textbf {\bibinfo {volume} {128}},\
  \bibinfo {pages} {140602} (\bibinfo {year} {2022})}\BibitemShut {NoStop}%
\bibitem [{\citenamefont {Allahverdyan}\ \emph {et~al.}(2004)\citenamefont
  {Allahverdyan}, \citenamefont {Balian},\ and\ \citenamefont
  {Nieuwenhuizen}}]{Allahverdyan_2004}%
  \BibitemOpen
  \bibfield  {author} {\bibinfo {author} {\bibfnamefont {A.~E.}\ \bibnamefont
  {Allahverdyan}}, \bibinfo {author} {\bibfnamefont {R.}~\bibnamefont
  {Balian}}, \ and\ \bibinfo {author} {\bibfnamefont {T.~M.}\ \bibnamefont
  {Nieuwenhuizen}},\ }\href {\doibase 10.1209/epl/i2004-10101-2} {\bibfield
  {journal} {\bibinfo  {journal} {Europhysics Letters ({EPL})}\ }\textbf
  {\bibinfo {volume} {67}},\ \bibinfo {pages} {565} (\bibinfo {year}
  {2004})}\BibitemShut {NoStop}%
\bibitem [{\citenamefont {Smolin}\ \emph {et~al.}(2012)\citenamefont {Smolin},
  \citenamefont {Gambetta},\ and\ \citenamefont {Smith}}]{smolin2012mle}%
  \BibitemOpen
  \bibfield  {author} {\bibinfo {author} {\bibfnamefont {J.~A.}\ \bibnamefont
  {Smolin}}, \bibinfo {author} {\bibfnamefont {J.~M.}\ \bibnamefont
  {Gambetta}}, \ and\ \bibinfo {author} {\bibfnamefont {G.}~\bibnamefont
  {Smith}},\ }\href {\doibase 10.1103/PhysRevLett.108.070502} {\bibfield
  {journal} {\bibinfo  {journal} {Phys. Rev. Lett.}\ }\textbf {\bibinfo
  {volume} {108}},\ \bibinfo {pages} {070502} (\bibinfo {year}
  {2012})}\BibitemShut {NoStop}%
\end{thebibliography}%

\end{document}